\documentclass[12pt]{article}

\usepackage{subfig} 
\usepackage{eufrak} 
\usepackage{amsmath}   
\usepackage[hyphens]{url}
\usepackage{hyperref}  

\usepackage{eurosym}   

\usepackage{mathptmx}
\usepackage{graphicx}
\usepackage[normalem]{ulem} 
\usepackage{color}

\usepackage{rotating}

\numberwithin{equation}{section}

\newcommand{\be}{\begin{equation}}
\newcommand{\ee}{\end{equation}}
\newcommand{\bea}{\begin{eqnarray}}
\newcommand{\eea}{\end{eqnarray}}

\newcommand{\sss}[1]{\scriptscriptstyle{#1}}
\newcommand{\zb}{Z}

\newcommand{\sman}{s}
\newcommand{\tman}{t}

\newcommand{\lpar}{\left(}                            
\newcommand{\rpar}{\right)}

\newcommand{\ib  }{i}
\newcommand{\qf  }{Q_f  }

\newcommand{\qe  }{Q_e  }

\newcommand{\vvertil}[3]{F^{#1}_{#2}\lpar{#3}\rpar}

\newcommand{\gadu}[1]{\gamma_{#1}}

\newcommand{\siws}{s^2_{\sss{W}}}

\newcommand{\tcie}{I^{(3)}_e}
\newcommand{\tcif}{I^{(3)}_f}

\newcommand{\gspi}{\frac{g^2}{16\pi^2}}

\def\gfd{\gamma_5}

    \newcommand{\stws}{s^2_{\sss{W}}}
    \newcommand{\stwf}{s^4_{\sss{W}}}

\title{THE ZFITTER PROJECT  
}

\author{\it
A. Akhundov$^{1}$,  
A.B. Arbuzov$^{2,3}$,
S. Riemann$^{4}$,  
T. Riemann$^{4}$
}

\date{}

\begin{document}

\maketitle

\begin{center}
{$^1$ 
Institute of Physics, Azerbaijan National Academy of Sciences, AZ-1143 Baku, Azerbaijan
} \\[.2cm]
{$^2$ 
Bogoliubov Laboratory of Theoretical Physics, Joint Institute for Nuclear Research JINR, 141980 Dubna,
 Russia
} \\[.2cm]
{$^3$ Department of Higher Mathematics, Dubna University, 141980 Dubna, Russia
} \\[.2cm]
{$^4$ 
15711 K{\"o}nigs Wusterhausen, Germany
}
\end{center}

\begin{abstract}
The ZFITTER project is aimed at the computation of high-precision theoretical predictions 
for various observables in high-energy electron-positron annihilation and other processes.
The stages of the project development are described.
Accent is made on applications to the analysis of LEP data. 
The present status of the project and perspectives are given as well. 
\end{abstract}

\clearpage

{\small 
\tableofcontents


\listoffigures
\listoftables
}

\clearpage

\section{Introduction}
\allowdisplaybreaks

\begin{figure}[t]
\hspace*{-1.0cm}
\includegraphics[width=9cm]{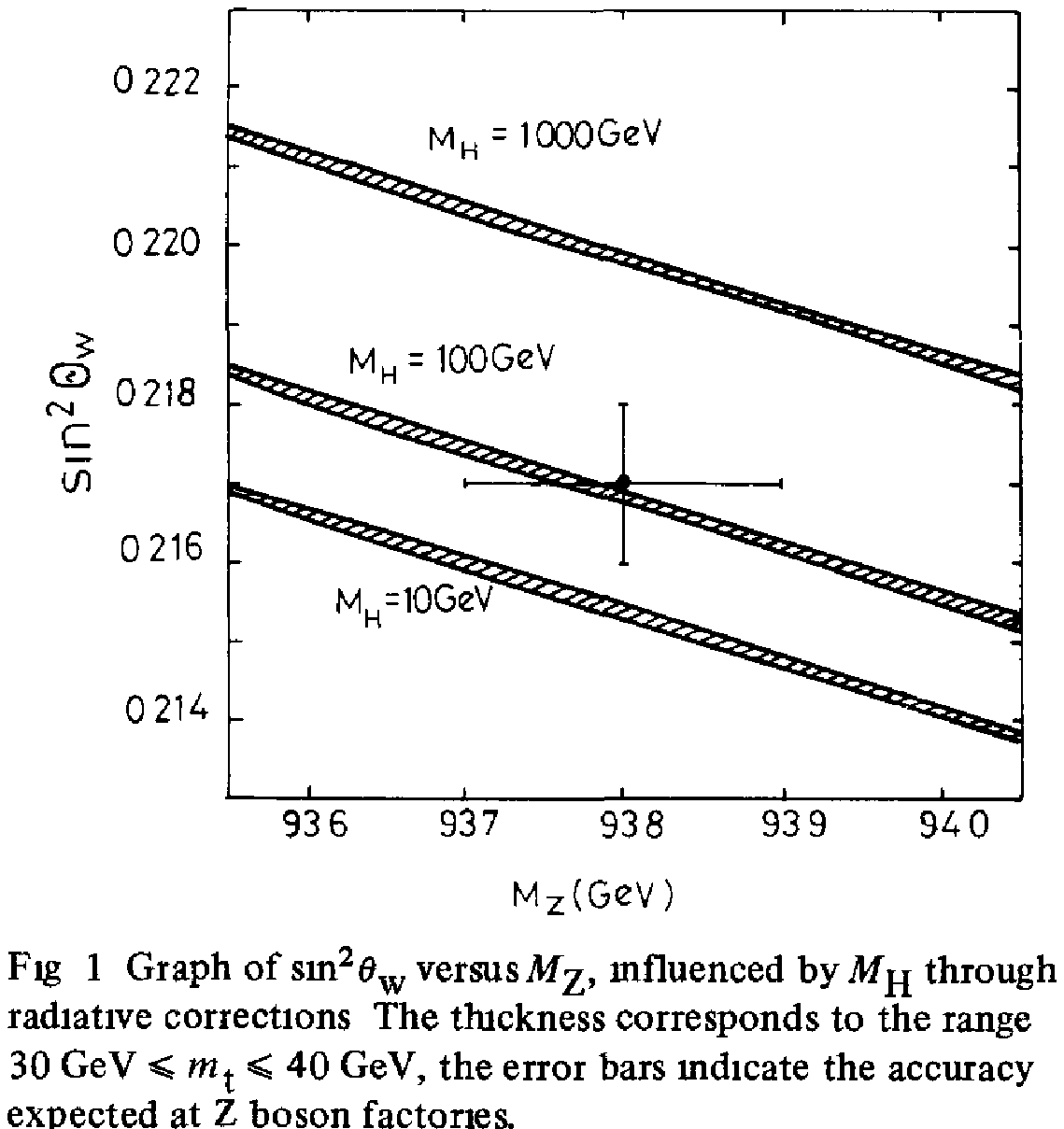} 
\includegraphics[angle=0,width=6.cm]{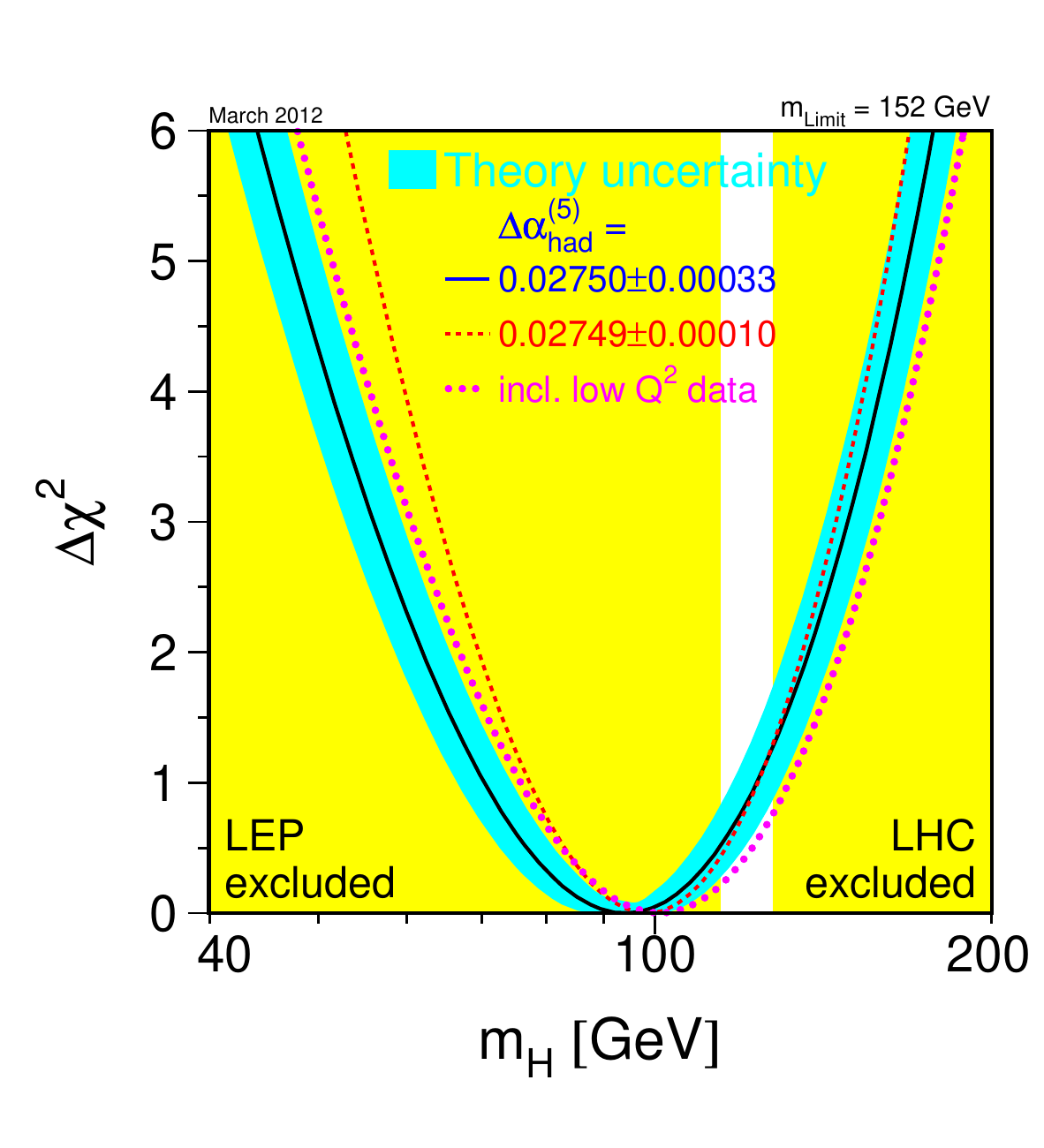}
\caption[Higgs mass and LEP measurements]{
\label{plb166fig1}
\textit{
Left: The first ever plotted LEP observables' dependence on the Higgs mass in
the Standard Model (reprinted from Physics Letters, A.~Akhundov, D.~Bardin, and T.~Riemann, ``Hunting the
hidden standard Higgs'', volume B166, p. 111, Copyright (1986) \cite{Akhundov:1985cf}, with permission from
Elsevier.)
Right: 
Blue-band plot of the LEPEWWG
\cite{lepewwg-webpage-jan-2013} with a Standard Model Higgs boson mass prediction based on combined world
data from precision electroweak measurements.
}}
\end{figure}

To name a date of begin of the ZFITTER project is difficult.
The first papers on electroweak loop calculations by D.~Bardin and O.~Fedorenko date back to 1976, 
but in another context.
In September 1983 the Dubna-Zeuthen group started activity, due to the begin of the four-year long stay of S.
Riemann and T. Riemann at JINR, Dubna.
The name ZFITTER was invented in 1989 and replaced the former name ZBIZON for the software project.
Finally, we chose the year 1985, when the article ``Hunting the hidden standard Higgs'' was published
\cite{Akhundov:1985cf}.
With this study, we began to take into account a finite, non-zero top quark mass $m_t$ in the radiative
corrections, in the context of $e^+e^-$-annihilation.
To our knowledge, the paper contains the first plot confronting two LEP observables -- weak mixing angle
$\sin^2\theta_W$ and  $Z$ boson mass $M_Z$ -- with their dependence on the unknown top quark mass $m_t$ and
the also unknown Higgs boson mass $M_H$ in the Standard
Model \cite{Glashow:1961tr,PhysRevLett.19.1264,Salam:1968rm,'tHooft:1972fi}. 
We reproduce the plot here in figure \ref{plb166fig1}, left.
Both top quark and Higgs boson were not yet discovered at that time, and the actual experimental values for 
$M_Z$ and $\sin^2\theta_W$ had  too huge errors to be included into the plot \cite{Wohl:1984kn}: $M_Z = 92.9
\pm 1 6$ GeV and $\sin^2\theta_W = 0.23 \pm 0.015$.
The numbers in the figure are based on the one-loop Standard Model prediction for $\Delta r$, the weak
correction to $G_{\mu}$, deserving few lines of Fortran code.
We remark as a curiosity that from 1985 to 2011, the article was quoted only once (by authors outside our
group).

The LEP/SLC collaborations made exciting measurements of the $Z$ boson resonance and of its mass, width, weak
mixing angle etc.,  with an unexpected final accuracy \cite{Beringer:1900zz}:
\bea
M_Z &= &91.1876 \pm 0.0021 {\rm~~GeV},
\\
\Gamma_Z  &=& 2.4952 \pm 0.0023 {\rm~~GeV},
\\
\sin^2 \theta_{\rm weak} &=& 0.22296 \pm 0.00028 ,
\\
\label{sinw2-lep}
\sin^2 \theta_{\rm lept}^{\rm eff} &=& 0.23146 \pm  0.00012,
\\
\sin^2 \theta_{Z}^{\rm MS} &=&  0.23116 \pm 0.00012,
\\
 N_\nu &=& 2.989 \pm 0.007.
\eea
For the $Z$ boson mass, this implies $\Delta M_Z/M_Z \approx 10^{-5}$.
For the various definitions of the  weak mixing angle see section 10 of \cite{Beringer:1900zz}.
And $N_\nu$ is the number of light neutrinos.

\begin{figure}
\begin{center}
\includegraphics[width=7.5cm]{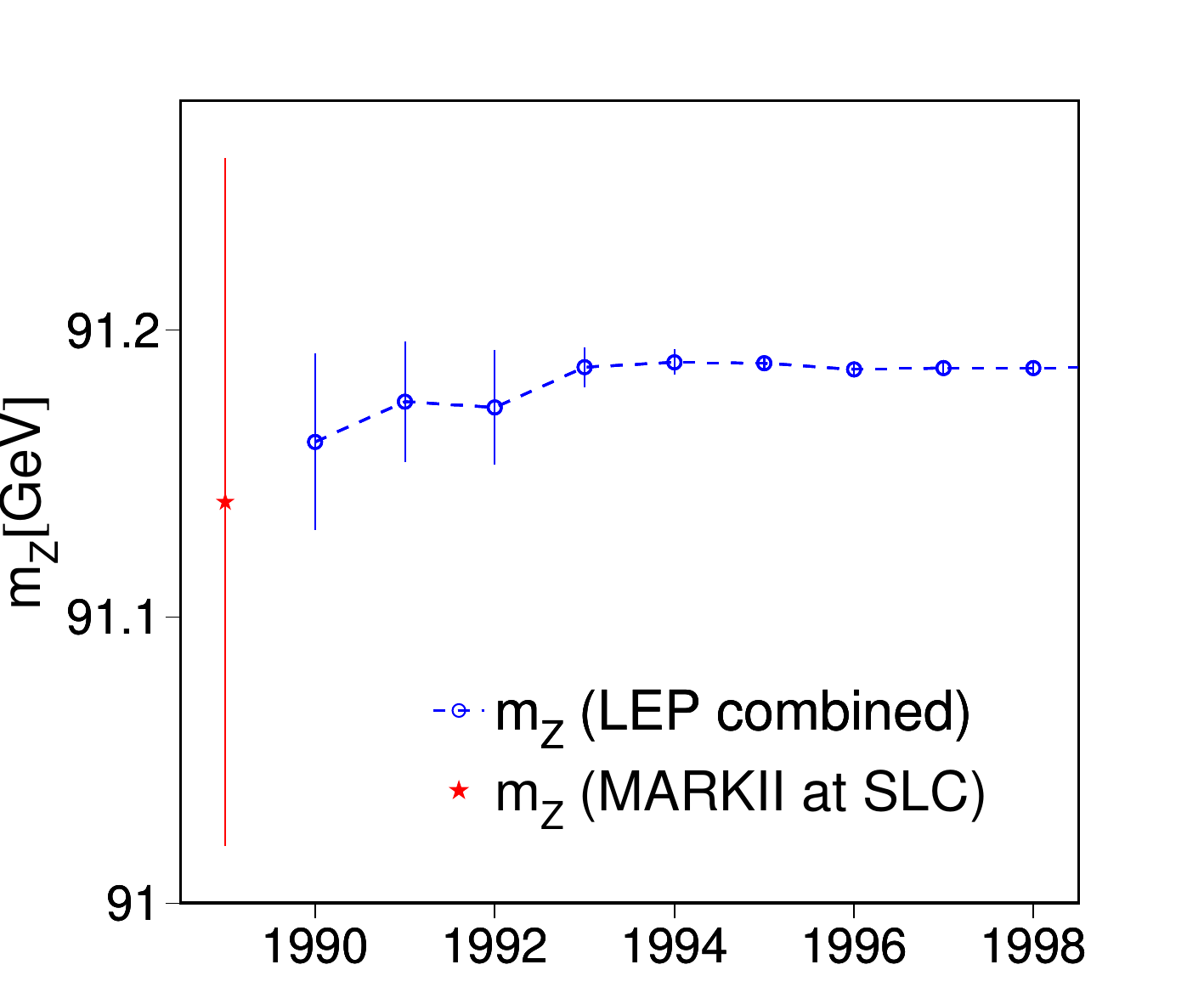}
\includegraphics[width=7.5cm]{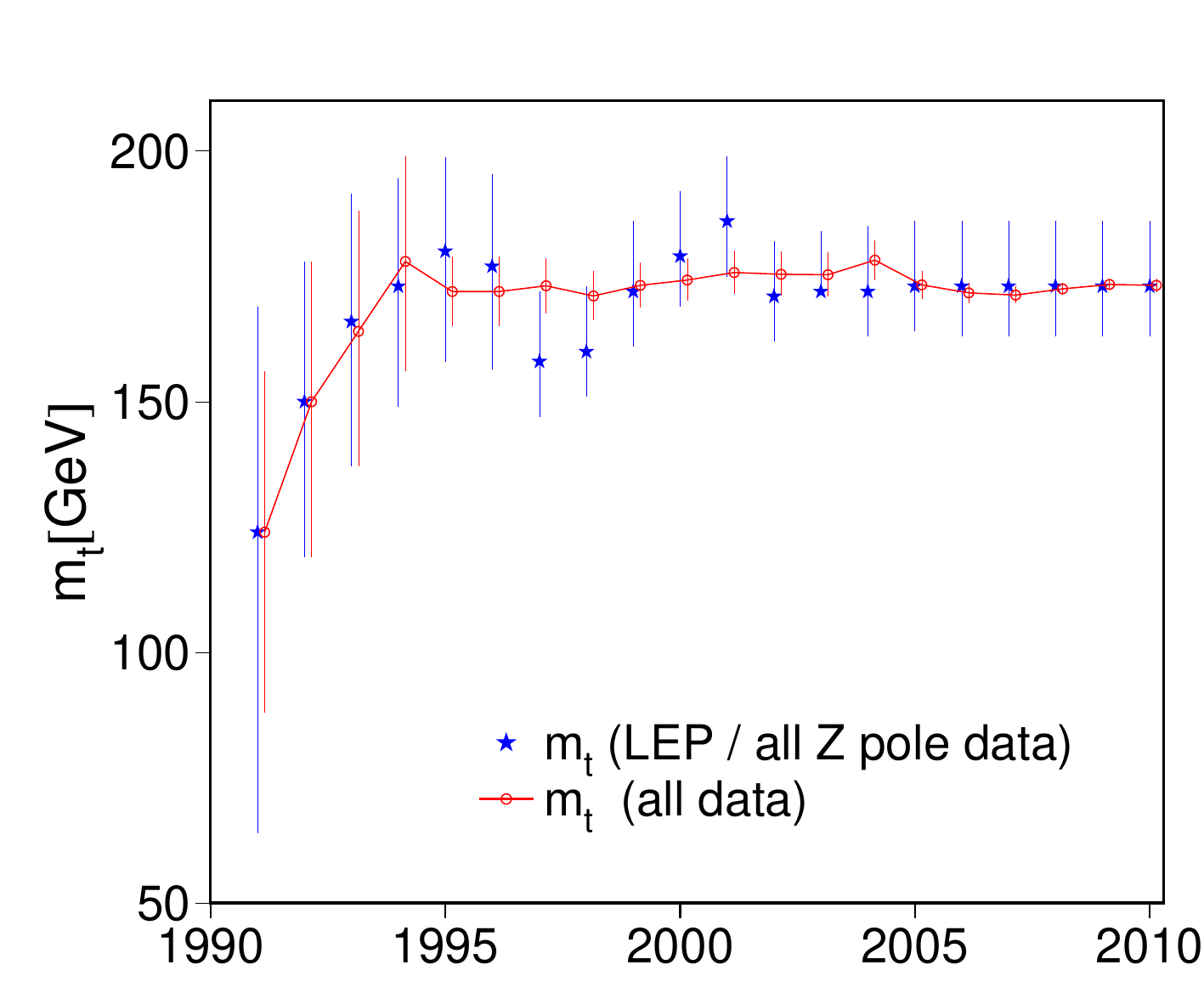}
\end{center}
 \caption[Z boson and top mass measurements]{\textit{
Left:
$Z$ boson  mass measurements at LEP. Earlier measurements are from UA1, UA2 at SPS (CERN) (see
text, not shown in plot) and from MARKII at SLC (SLAC).
Right:
Top quark mass measurements.
} 
\label{fz}}
\end{figure}

\begin{figure}
\begin{center}
\includegraphics[width=7.5cm]{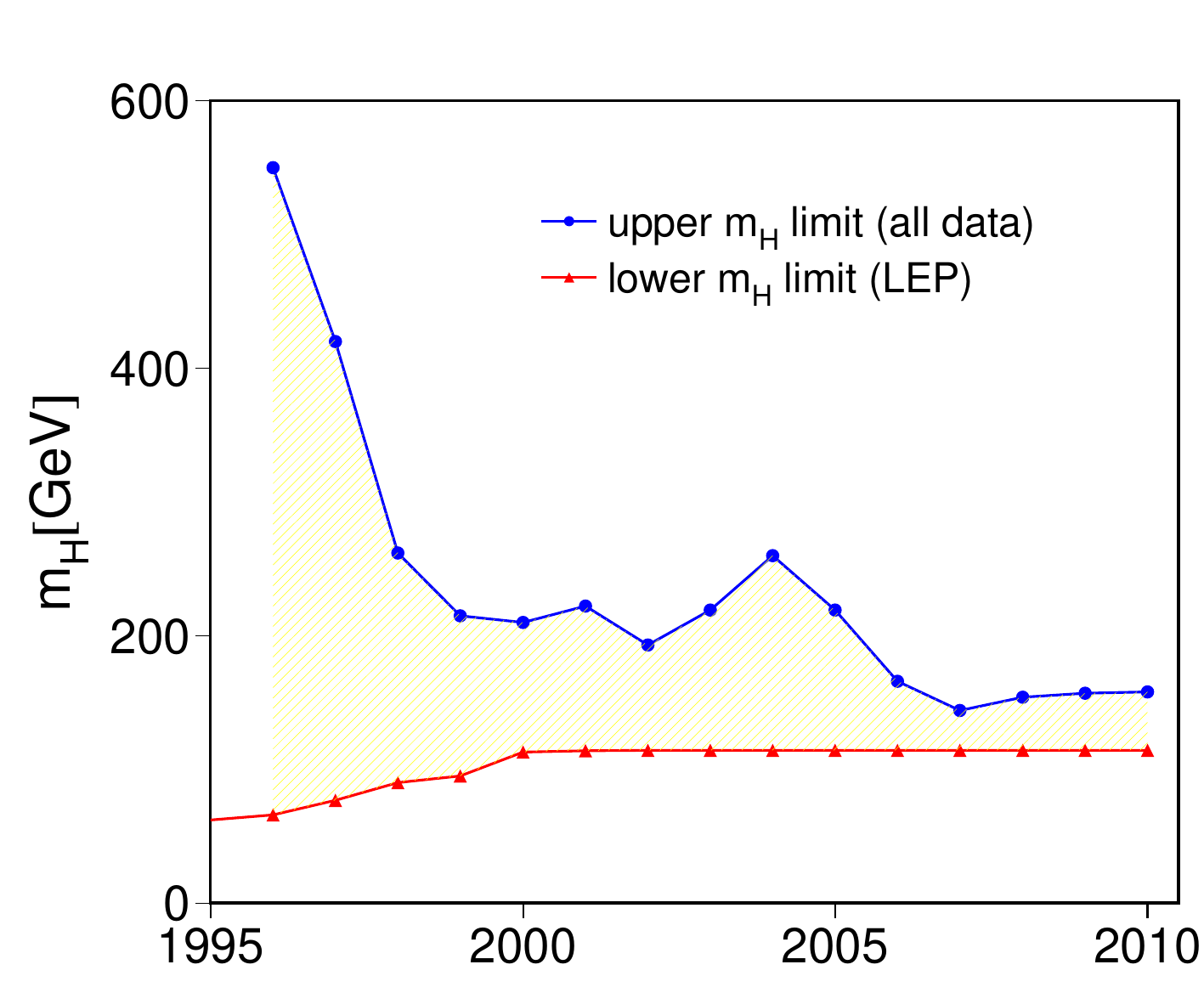}
\end{center}
\caption[Higgs boson  mass measurements]{
\textit{
Higgs boson  mass measurements. The upper limits and the fit values for $M_H$ derive from a
combination of virtual corrections to LEP and similar data, top and $W$ mass measurements, performed by the
LEPEWWG. The lower mass limit is due to LEP direct searches. The lower limits from data combinations are not
shown.
  \label{fh}}
}
\end{figure}

Figure \ref{fz} shows the rise of accuracy for $M_Z$ due to LEP.
Since the begin of the nineteen-nineties, a true scientific standard is the so-called blue-band plot of
the
{LEPEWWG}\footnote{\url{http://lepewwg.web.cern.ch/LEPEWWG/}}, based on ZFITTER
\cite{Bardin:1989di,Bardin:1989tq,Bardin:1992jc,Bardin:1999yda,Arbuzov:2005ma} and another Standard Model
package TOPAZ0
\cite{Montagna:1993ai,Montagna:1995ja,Montagna:1998kp}.\footnote{Reference \cite{Bardin:1992jc} (1992)
appeared as a CERN preprint because, at that time, we considered this to be more prestigious than e.g. a paper
in the journal ``Computer Physics Communications'' devoted to publication of software. It was submitted to the
Internet archive hep-ph in 1994.} 
The March 2012 version is reproduced in figure \ref{plb166fig1}, right.
Both ZFITTER and TOPAZ0 are huge software packages with tens of thousands lines of Fortran code aiming at
covering the
complete known radiative corrections to the $Z$ resonance peak in the reaction $e^+e^- \to {\bar f} f$.
The top quark was predicted by M. Kobayashi and T. Maskawa in 1973 \cite{Kobayashi:1973fv} and discovered in
1995 with a mass of about 173 GeV \cite{Abe:1995hr,Abachi:1995iq}.
Top quark mass data from precision electroweak measurements and from direct searches are collected in figure
\ref{fz}, right. 
 After discovery of the top quark, the LEP data were no more competitive.
The agreement of direct measurements (in `all data') and indirect measurements (in `all $Z$ pole
data') supports the validity of the Standard Model at the quantum
loop level.
Over the years the predictive power of the indirect searches for the Higgs boson mass improved
considerably, and the discovery of the top quark was a crucial input.  
This is described in figure \ref{fh}.
In 2012, the LHC collaborations reported the discovery of a scalar particle with a mass of about 125 GeV
\cite{atlas:2012gk,CMS:2012gu}, which fits into these expectations from the indirect searches.
The general believe is that this particle is (similar to) the one predicted by Peter Higgs in 1964
\cite{Englert:1964et,Higgs:1964pj,Higgs:1964ia}.
Within less than a year, in October 2013, Peter Higgs and Francois Englert were awarded the Nobel Prize
in Physics ``... for the theoretical discovery of a mechanism that contributes to our understanding of the
origin of mass of subatomic particles, and which recently was confirmed through the discovery of the predicted
fundamental particle, by the ATLAS and CMS experiments at CERN’s Large Hadron Collider''
\cite{NP2013physicspressrelease}.
The accompanying advanced public information  ``Scientific Background on the Nobel Prize in Physics 2013:
The BEH-Mechanism, Interactions with Short Range Forces and Scalar Particles'', compiled by the
Class for Physics of the Royal Swedish Academy of Sciences \cite{advancedphysicsprize2013}, reproduces the
Blue-band plot (March 2012) of the LEPEWWG on page 16. The plot relies on ZFITTER v.6.43.

We live with the ZFITTER project for about 30 years now,
and ZFITTER is yet in use for a diverse
variety of
applications, ranging from the global analyzes of the LEPEWWG to many graduation papers, habilitations
like e.g. \cite{Mnich:1996hy} and PhD theses like \cite{Eberhardt:2013wia}.
Thirty years are a long term.
It takes similarly long to prepare final results of big experiments at accelerators  as
LEP~1 (running 13 August 1989 - 1995), LEP~2 (running 1996 - 2 November 2000), HERA (running
1992-2007).
The final analysis of the LEP~1 data for two-fermion production was published in 2005
\cite{ALEPH:2005ab} by the LEP collaborations and the LEPEWWG, using ZFITTER v.6.42.
The corresponding enterprise for LEP~2 data was finalized these days \cite{Schael:2013ita}, using
ZFITTER
v.6.43.

The big laboratories invented scientific programs for a dedicated long-term preservation of
the experimental data, under the label ``ICFA Study Group on Data Preservation and Long Term Analysis in High
Energy Physics'' \cite{dphep:2013}. 
One might assume that this is a self-evident issue of any physics collaboration. 
Physics is the science of
reproducible observations in Nature and of their explanations/descriptions, and reproducibility deserves
storage.
But long-term storage is an unsolved problem, worth of any
(reasonable) effort.
DESY, as an example, founded in 2009 a ``DESY Data Preservation Project'', mainly focusing on the HERA
experiments H1, Hermes, ZEUS \cite{dpdesy:2012}.
If such effort is justified for data, then it is also needed for the analysis tools, which were
used for an extraction of the Model with its few parameters from the raw, or not-so-raw, Data.
To our knowledge, the Big Labs do not plan to support long-term maintenance of software like ZFITTER.
We, as the authors, theoreticians and phenomenologists, have to mind by ourselves about
maintenance of theory/phenomenology software.
Everybody knows that the very details of a data analysis cannot be described by few words.
But for precision studies they are truly essential.
Sometimes we say: ``The description of the program is the program
itself.''
This is a helpful statement if ``the program itself'' is preserved over a long term in its state of use.
ZFITTER did and does a lot to fulfill such a demand.
See the web-page \url{http://zfitter.education}.

Preservation demands effort. 
There are 17 people involved in the DESY Data Preservation Project. 
At the other hand, if a theoretician says: I care about the availability of my old software, people
start to smile. 
This aim does not give true credit points for a scientific carrier, in what phase of the
carrier ever.
In fact, not only the so-called main author of ZFITTER, D.~Bardin, our
{\it ``primus inter pares''}, tends to lose interest in active support of ZFITTER over the decades. 
This applies to all of us, mainly because of our interest in studying or inventing something new.
Nevertheless, we collected in 2005 some volunteers into a ZFITTER support group, which submitted in that
year ZFITTER v.6.42 and in 2008 ZFITTER v.6.43 \cite{Bardin:1999yda,Arbuzov:2005ma}.
The ZFITTER v.6.44beta version dates in 2013 \cite{ZFITTERv644beta:2013}.
Encouraged by the decreasing visibility of our ZFITTER support, in 2006 some experimentalists tried to
re-program in C++ in a year's time the Standard Model library of ZFITTER from the published literature.
Not just for fun, but in order to do better than ZFITTER: use a more modern programming language than
Fortran, with more modularity than ZFITTER, a bit updated, with a GUI. In order to retain ZFITTER for a longer
term.
The project was proprietary until August 2012, and it faced two major problems.
It proved to be impossible to do so without using the ZFITTER software itself to a large extent.
Further, without cooperation with ZFITTER authors and the community of theoreticians, including extensive
numerical cross-checks, such a project cannot succeed.\footnote{See subsection \ref{sec-gfitter}.}

Finally, there is much influence by institutes' directors and by the  editors and publishers of physics
journals on the engagement of scientists in the development of software.
Not all of them seem to mind about proper acknowledgment and quotation of software.  
Some even say that software has no genuine scientific value by itself and advocate an absolutely free use of
any software as common habit.
If this {\em would become} common habit,
nobody with inspiration and ambition would invest time to write complicated
software
for the use by other people, like the ZFITTER group - and other groups as well - does.
We live in an academic world and we are valued by our scientific results, their originality,
importance,
curiosity, usefulness etc.
Financing of our projects, of our working positions, our academic prestige depend on all that.  
We need proper quotation of our scientific results in case they are used.
And we can only appeal and hope that the community understands this as a justified expectation, also for
software.\footnote{Note added in May 2014:
See the summary of a round table discussion at ACAT2013 \\
\url{http://acat2013.ihep.ac.cn/proceedings/papers/D002-142-Round_Table_Open-source,_knowledge_sharing_and_scientific_collaboration.pdf},
.}
As a key feature of user-friendly support, we stored for many years all the relevant versions of
ZFITTER at the project web-page for anonymous download. We collected about three dozen versions,
covering more than 20
years.
There are colleagues who take the freedom to use ZFITTER as if it were
{open-source software}\footnote{\url{http://en.wikipedia.org/wiki/Open_source_software}}
in the strictest meaning of the word.
Despite the facts that academic research deserves strict, proper quotation, and that there are license
regulations (for ZFITTER this includes the 
{CPC license}\footnote{\url{http://cpc.cs.qub.ac.uk/licence/licence.html}}). In some countries there are even
legal
regulations.\footnote{Due to controversial positions, we closed the links for anonymous download from
ZFITTER web-pages in 2011; in 2012 the copies in the Andrew file system at CERN were removed.} 

It is the aim of these notes to give an overview on the ZFITTER project.
Maybe they can help to see theoretical software in particle physics as an intellectual enterprise like
the other inventions of physics research - experimental set-ups, data, hypotheses, models, theories.

\medskip

We would like to finish the introduction with two quotes.

Several times we all thought that the ZFITTER project is in its final phase of dying out.
See for example the  remark of Dima Bardin at the symposium ``50 Years of Electroweak Physics: a
symposium in honor of Professor Alberto Sirlin's 70th
Birthday'', in the year 2000 \cite{Bardin:2001sk}:

\begin{footnotesize}
\textit{``We would like to see the end of the ZFITTER project in the year 2000 and, therefore, a very natural
question
arises: What's next?''}
\end{footnotesize}

In the same year, members of the ZFITTER group 
were granted the prestigious  Award in Theoretical
Physics of the Joint Institute for Nuclear Research, Dubna, Russia.
\href{http://zfitter.education/jinr-prize-certificate-t-riemann-2.pdf}{certificate}.
The referee was Academician Prof. L.B.~Okun from ITEP Moscow; he finished his estimate with the
statement:\footnote{The
original document is in Russian, see \url{http://zfitter.com/jinr-prize-okun.pdf}.}

\medskip

\begin{footnotesize}
\textit{``Overall, the project "ZFITTER Fortran program" represents a unique theoretical tool of
world class. The project
formed the basis of a close cooperation of experimentalists and theoreticians (with a series
of workshops at CERN). With the accumulation of 
experimental data, the accuracy of the programs has been increased. The project has always
found great interest at conferences. Its importance and the interest
to it shows with numerous references in articles, reviews and monographs.
In the long term, with the advent of more precise experiments, ZFITTER
will allow to take into account all two-loop electroweak corrections.
\\
The series of theoretical articles on precision tests of the Standard Model at electron-positron
colliders certainly deserves the award of the JINR prize 2000.
\\
Academician L.B. Okun''}
\end{footnotesize}
\bigskip

Our figures illustrate the development of mass predictions for $Z$ boson (figure \ref{fz}, left),
top quark (figure \ref{fz}, right), Higgs boson (figure \ref{fh}). 
Here, ZFITTER has been useful until now.
Okun's proposition that ZFITTER will be used also in future is being  fulfilled.
We can only hope that our write-up might help to convince the present particle 
physics community that ZFITTER is worth some support by now and in future.

At the end of the introduction, we would like to reproduce the long(est) authors list of ZFITTER, see also
\url{http://zfitter.education}:

A.~Akhundov, A.~Arbuzov, M.~Awramik, D.~Bardin, M.~Bilenky, A.~Chizhov, P.~Christova, M.~Czakon,
O.~Fedorenko (1951-1994), A.~Freitas, M.~Gr\"unewald, M.~Jack, L.~Kalinovskaya, A.~Olshevsky, S.~Riemann,
T.~Riemann, M.~Sachwitz, A.~Sazonov, Yu.~Sedykh, I.~Sheer, L.~Vertogradov, H.~Vogt.

The list is not complete.
According to the conventions of the software library of ``Computer Physics Communications'', we should also
include here all the co-authors who helped to prepare the program descriptions in 1989, 1999, 2005
\cite{Bardin:1989tq,Bardin:1999yda,Arbuzov:2005ma}.

\section{ZFITTER in a nutshell, or: Is there a ZFITTER approach?}

We never used the label ``ZFITTER approach''.
The reason is simple: There is no ZFITTER approach.
If any, there is a kind of Dubna approach, or of Bardin's group's approach.

Nevertheless, other people use this phrase.
Let us collect some distinguishing moments which might be the origin of some popularity
of ZFITTER, but also of one or the other of our scientific projects:

\begin{itemize}
     \item  
Unitary gauge.
\\
We are working in the unitary gauge  when studying the renormalization of the Standard Model.
Most of the other groups use the 't Hooft-Feynman gauge.
But when looking at observable quantities, there is no difference left, due to the gauge invariance of
perturbation theory. So, if everything is correct, there is no difference for the users.       
\item
On-mass-shell renormalization scheme.
\\
We are  applying the on-mass-shell renormalization scheme, with few modifications.
Other groups do the same for electroweak corrections.
    \item 
Analytical treatment of
QED corrections.
\\
\emph{ZFITTER is not a Monte-Carlo program.}
The Dubna group has an enormous experience in the analytical treatment of
QED corrections, allowing us, sometimes,  to come relatively close to the experimental set-ups by dedicated
analytical
integrations. Several different approaches may be chosen by users.
The necessary computational time for fits to data is small compared to that of other projects.
   \item  
Realistic observables and pseudo-observables.
\\
There is a plethora of observables, of quite different polarized and non-polarized cross-section
combinations
and asymmetries.
Both so-called realistic observables (including real corrections) and pseudo-observables (after unfolding the 
realistic observables) may be used.
With the different interfaces one may optimize a study appropriately.
\item
Form factors. Modularity.
\\
We describe the effective Born cross-section in the Standard Model approach by (essentially) four
(complex-valued) gauge invariant form factors per production channel.\footnote{Massive top quark production
deserves six form factors \cite{Bardin:2000kn,Fleischer:2002rn,Fleischer:2003kk}. See also the in-depth
discussion in \cite{Beenakker:1989}.}
Plus a separated running QED coupling.
This allows a modular programming, the efficient introduction of New Physics into the package, or the
convenient export of the Standard Model corrections into another approach to the real corrections. 
\item
Higher-order corrections. 
\\
Originally, we calculated the complete electroweak one-loop corrections to the $Z$ resonance physics.
By time, there became  more and more electroweak, QCD, and mixed  higher-order corrections available, 
and we had to implement them into ZFITTER.
In the nineteen-nineties these implementations dominated our efforts for ZFITTER.
It is not the genuine theoretical work we like, but has to be done.
    \item
Interfaces. Modularity.
\\
\emph{ZFITTER is not a fitting program. }
But from the very beginning, we were aware of the fact that a data analysis at e.g. LEP may rest on
different sets of  assumptions, being incompatible to each other.
The notion of interfaces was developed.
The interfaces call the kernel of ZFITTER with different compositions of input variables, real corrections, an
effective Born cross section. 
The users of ZFITTER can choose among few sample interfaces, or they write their own ones.
\item
Flags.
\\
The use of ZFITTER may be controlled by flags to be set by users.
Although this implies problems for the update by the authors, for users this is truly convenient.
\item
Descriptions.
\\
ZFITTER is described for users at different levels of complexity.
There are about 350 pages of instructions.
\item
Simplicity of file structure.
\\
ZFITTER is 
easy to use.
It has a simple file structure, is self-contained, and has a sample output.
The installation at a computer is done and controlled within minutes.
The installation of the user software, which is calling ZFITTER and performing data fits, writing
tables and
drawing figures, might be much more involved.
\item
Numerical cross-checks.
\\
With very precise data available, as it was typical for LEP physics, a careful numerical control of the
theory software became mandatory.
Here, a lot of colleagues, including competitors of ZFITTER, invested huge collective efforts.
Without that, one could neither trust the impressive physical results of that era nor the
long-term reliability
of the code.
\item
Source-open programming.
\\
The scientific seriosity of ZFITTER is trustable because its source code is publicly available.
Because we expect that the usual academic conditions of use are respected, notably the CPC license, we say it
is source-open software. 
The meaning of the word open-source software is controversial and it should not be used for ZFITTER.  
\item
Social aspects.
\\
A software package of some complexity, written for use by other people, must be supported and, in case,
updated.
The authors need some contact with the users.
And, last but not least, some license regulations have to be fixed if the authors want to get their academic
credit, e.g. in form of proper citations.
Since the authors of ZFITTER are employed at some institutions distributed over several countries, it is
of vital importance that these institutions interfere in a constructive way. 
We are happy that this did happen for a very long period, in view of several social restructurings of
institutions and even countries. 
\end{itemize}

ZFITTER is a Fortran library of Standard Model predictions for the scattering process
\bea
e^+e^- \to {\bar f} f ~~(+ \gamma, ~~ + n~\gamma)
\eea
at energies in the range  {  $\sqrt{s} \approx$ 20 GeV to 150 GeV},
above quark bound states [meson factories] and below the top threshold.
The package is to be called by {  interfaces}
\begin{itemize}
 \item in the {Standard Model};
\item in several  {model-independent approaches};
\item with {$Z'$ bosons} and similar physics extensions;
\item etc.
\end{itemize}
One may evaluate 
\begin{itemize}
   \item realistic observables -- polarized and non-polarized cross-sections and cross-section asymmetries
with a variety of cuts on the final state 
\item (pseudo-)observables like ~~~ $M_Z, ~~~  \Gamma_Z,  ~~~ \sigma_{\rm had}^{\rm tot}, ~~~ R_{\rm had}, ~~~
A_{\rm FB}^{\rm lept}, ~~~ \lambda_{\tau},~~~ \sin^2 \theta_{\rm ew}^{\rm eff}, ~~~\cdots$
\item 
the form factors, for use in another analysis program.                 
\end{itemize}
with different choices of input variables, e.g.
\begin{itemize}
\item
$ M_Z, ~~~G_{\mu}, ~~~m_{t}, ~~~M_H,~~~ \alpha_{\rm em}, ~~~ \alpha_{s}, ~~~\cdots$
   \item 
 $M_Z, ~~~M_W, ~~~m_{t}, ~~~M_H,~~~ \alpha_{\rm em}, ~~~ \alpha_{s}, ~~~\cdots$
  \end{itemize}

\allowdisplaybreaks

\section{Electroweak virtual corrections}

The first weak one-loop calculations  were published as Dubna preprints by D.~Bardin and his PhD student 
O.~Fedorenko in 1978 \cite{Bardin:30845,Bardin:30846,Bardin:11461}.
Together with  P.~Christova, then also PhD student of D.~Bardin, the by now famous articles on the
complete
on-mass-shell renormalization of the electroweak Standard Model were published in Nuclear Physics B
\cite{Bardin:1980fe,Bardin:1981sv}, for fermion scattering. See also reference \cite{Bardin:1982ev}. 
The corresponding studies for weak boson production and fermion--boson scattering are  unpublished
\cite{Bardin:1982xv,Bardin:1982my}.

These calculations were complete, but assumed all fermions to be massless.
When experiments showed that at least the top quark should be heavy, the top mass dependence was included
\cite{Akhundov:1985cf,Akhundov:1985fc,Bardin:1986fi,Bardin:1989di}.
Some studies of structural aspects in the renormalization of the Standard Model are
\cite{Khristova:1987pd,Czyz:1987xx}.
All this was done in the unitary gauge, while most of other groups usually worked with the
't~Hooft-Feynman
gauge. Later, this difference was of some value because an agreement of two calculations performed in truly
quite different gauges establishes a powerful cross-check of the numerics.
The first numerical program BFK (acronym for Bardin/Fedorenko/Khristova) was written in Fortran.

The Zeuthen partners, staying at Dubna from 1983 to 1987, worked out the renormalization of the electroweak
Standard Model in the 't Hooft-Feynman gauge \cite{Mann:1983di}.
But because there was never a numerical program created, the results of this work were more or less useless;
they had a mere educational value.
Nevertheless, the experiences from that activity were used in order to perform the first calculation of
flavor-changing $Z$
boson decays into different lepton flavors.\footnote{%
We mention for curiosity that the numerics of this one-loop project was performed with a pocket calculator
TI-57 with 50 program steps. The program had to be typed in after switching on. The price of the device was
120 DM in the CERN shop.}
 This was unpublished \cite{Mann:1981hq,Mann:1982du}; see also
\cite{Riemann:1982rq}.
An application to flavor-violating $Z$ decays into different quark flavors was finally published
\cite{Mann:1983dv}.
Later, when we were working on precision predictions for LEP, the results could be easily transformed 
into the
calculation of virtual top mass corrections in (flavor-diagonal) $b {\bar b}$ production at LEP and
in $Z$ decay \cite{Akhundov:1985fc}. 
And yet later, they were a starting point for studies of lepton number violation in $e^+e^-$ annihilation
with heavy neutrinos \cite{Illana:2000ic} and with supersymmetry \cite{Illana:2002tg}.\footnote{Several of the
results in supersymmetry found in the literature turned out to be just wrong when we had a look at them.}

\subsection{Sirlin's approach}
The notion of form factors $\rho$ and $\kappa$ in the weak neutral current were, to our knowledge, introduced
by {  A. Sirlin}:\footnote{For a historical perspective, see  reference  \cite{Sirlin:2012mh}.}
\begin{itemize}
 \item  $\rho$ -- contains the electroweak corrections to the Fermi constant $G_{\mu}$;
 \item $\kappa$ -- contains the electroweak corrections to the weak mixing angle $\sin^2\theta_{W}$.
\end{itemize}
This approach allows to retain in the on-mass-shell renormalization scheme the Born definitions
also in higher orders:
\bea
G_F^{\rm eff} &=& {\rho_Z} ~ G_{\mu},
\\
\sin^2 \theta_{W}^{\rm eff} &=& {\kappa_Z} ~ \sin^2\theta_{W},
\eea
where
\bea
\frac{G_\mu}{\sqrt{2}} &\equiv& \frac{g^2}{8M_W^2},
\\
\sin^2\theta_{W} &\equiv& 1- \frac{M_W^2}{M_Z^2}.
\eea

\subsection{The HECTOR and ZFITTER approach \label{sub-ff}
}
For general 4-fermion scattering amplitudes, one needs a more general
description.
This was first introduced, to our knowledge, by the Dubna/Zeuthen group, in 1987/88, in the article  
{``Electroweak Radiative Corrections to Deep Inelastic Scattering at HERA. Neutral Current
Scattering''}
by 
D.~Bardin, C.~Burdik (Dubna), P.~Khristova (Shoumen), T.~Riemann (Zeuthen)
 \cite{Bardin:1987rz,Bardin:1988by}.
The corresponding software is retained until today as the Fortran package  HECTOR \cite{Arbuzov:1995id}.
So, strictly speaking, one might call this the HECTOR approach.
 
We use four complex form factors $\rho, \kappa_{\rm ini}, \kappa_{\rm fin}, \kappa_{\rm ini-fin}$ for the
parameterization of the
weak amplitude, {including the $WW$ and $ZZ$ box diagrams}.
In the article  	
{``A Realistic Approach to the Standard Z Peak''} by 
D.~Bardin, M.~Bilenky, G.~Mitselmakher (Dubna), T.~Riemann, M.~Sachwitz (Zeuthen) \cite{Bardin:1989di},
we {excluded the weak $WW$ and $ZZ$ box} diagrams from the form factors, making them {independent
of the scattering angle}.
This is of advantage at LEP where these box diagrams have minor numerical influence.
When form factors are independent of the scattering angle, analytical phase space integrations become
possible.
In ZFITTER, there is an option to switch between the approaches. 

The Born amplitude is factorized into two pieces with vector coupling $v_i$ and axial vector coupling $a_i$ of
a fermion $i$ to the $Z$-boson; with $A_i = \gamma_{\mu}(v_i+a_i\gamma_5)$:
\bea  
A_i \otimes A_f \equiv \left[ {\bar u_i} \gamma_\mu (v_i+a_i\gamma_5) u_i \right] \times  
\left[ {\bar u_f} \gamma^\mu(v_f+a_f\gamma_5) u_f \right].
\eea
This form is generalized by loop corrections to
\bea  
A_{vv} \gamma \otimes \gamma + A_{av} \gamma \gamma_5 \otimes \gamma + A_{va} \gamma \otimes \gamma \gamma_5 +
A_{aa} 
\gamma  \gamma_5 \otimes \gamma \gamma_5 ,
\eea 
or, equivalently,
\bea  
B_{LL} \gamma (1+\gamma_5) \otimes \gamma(1+\gamma_5) + B_{\gamma L} \gamma  \otimes \gamma(1+\gamma_5) 
+ B_{L\gamma} \gamma(1+\gamma_5) \otimes \gamma + B_{\gamma\gamma} \gamma \otimes \gamma .
\eea 
{With $Z$ boson and photon exchanges:}
\bea 
{\cal M} &=&  {\cal M}_{\gamma}+{\cal M}_Z,
\\ 
{\cal M}_{\gamma} &\sim& {F_A} \left[\gamma \otimes  \gamma \right],
\\ 
{\cal M}_Z &\sim& G_\mu ~  {\rho_Z} \left[ 
\gamma \gamma_5 \otimes  \gamma \gamma_5
+ {v_q}
\gamma \otimes  \gamma \gamma_5
+ {v_l}
\gamma  \gamma_5 \otimes  \gamma
+ {v_{ql}}
\gamma \otimes  \gamma
\right].
\eea
In Born approximation, it is
\bea
 {v_{ql} \approx v_q \times v_l} .
\eea
{The form factors ~~$F_A,~~ \rho,~~  \kappa_q,~~  \kappa_l,~~  \kappa_{ql}$} are complex-valued functions of
$s$ and $t$:
\bea 
{F_A(s)} &=& \frac{\alpha_{QED}(s)}{\alpha_{em}}
\\\nonumber
 &=& 1 + \delta\alpha_{QED}(s),~~~
\\
\alpha_{em} &=& \frac{1}{137\cdots},
\\ 
a_f&\equiv& 1, ~~~~~~~~~~~~~~~~~~~~~~~~~~~~~~~~~~~~~~~~~f=q,l
\\ 
v_f(s,t)^{\rm eff} &=& 1-4\sin^2\theta_w |Q_f|{\kappa_{f}(s,t)}, ~~~~f=q,l
\\ 
v_{ql}(s,t)^{\rm eff} &=&v_q+v_l-1 +16\sin^4\theta_W |Q_qQ_l|{\kappa_{ql}(s,t)} ,
\eea
where we use $Q_e=-1$.
{From \cite{Bardin:1999yd1}, eq. (3.3.1), we quote:}
\begin{eqnarray}  
{\cal A}^{\sss{OLA}}_{\sss{\zb}}(\sman,\tman)&=& 
        \ib\,e^2\,4\,\tcie\tcif\frac{\chi_{\sss{Z}}(\sman)}{\sman}
        {\rho_{ef}(\sman,\tman)}
        \biggl\{
        \gadu{\mu}{\lpar 1+\gfd \rpar }
        \otimes \gadu{\mu} { \lpar 1+\gfd \rpar}    
              \\ \nonumber            &&
-4 |\qe | \stws {\kappa_e(\sman,\tman)}
        \gadu{\mu} \otimes \gadu{\mu}{\lpar1+\gfd\rpar}
       -4 |\qf | \stws {\kappa_f(\sman,\tman)}
        \gadu{\mu} {\lpar 1+\gfd \rpar } 
        \otimes \gadu{\mu}                              
             \\ \nonumber      &&
+
16 |\qe \qf| \stwf {\kappa_{ef}(\sman,\tman)}
        \gadu{\mu} \otimes \gadu{\mu} \biggr\}.
\label{processrhokappadef}
\end{eqnarray}
The form factors may be used, in analogy to the $Z$ decay matrix element of Sirlin, for  definitions of
effective vector and axial vector couplings and of a generalization of the effective weak mixing angle:
\bea
G_{\mu}^{\rm eff} &=& \rho_{ef} G_{\mu},
\\
\sin^2 \theta_{W,e}^{\rm eff} &=& \kappa_e ~ \sin^2 \theta_{W},
\\
\sin^2 \theta_{W,f}^{\rm eff} &=& \kappa_f ~ \sin^2 \theta_{W},
\\
\sin^2 \theta_{W,ef}^{\rm eff} &=& \sqrt{\kappa_{ef}} ~ \sin^2 \theta_{W}.
\eea
The unique definition of an effective weak mixing angle is lost.

The first applications of the calculations of weak corrections by the Dubna group were applied,
together with N. Shumeiko, to deep-inelastic scattering; see e.g. 
\cite{Bardin:1978gd,Bardin:1979qy}, based on the close relations to the NA-4 experiment at
CERN with JINR participation.
The form factors $\rho$ and $\kappa$ are simply related to
the one-loop
form factors introduced in the original renormalization articles by  Bardin and  Fedorenko (1978)
\cite{Bardin:30845,Bardin:30846,Bardin:11461} and 
Bardin, Christova, Fedorenko (1980) \cite{Bardin:1980fe,Bardin:1981sv}:
\begin{eqnarray}  
\label{form1}
\rho_{ef}  &=& 1+\vvertil{}{\sss{LL}}{\sman,\tman}-\siws\Delta r,       
\\  
\kappa_{e} &=& 1+\vvertil{}{\sss{QL}}{\sman,\tman}
                -\vvertil{}{\sss{LL}}{\sman,\tman},  
\\  
\kappa_{f} &=& 1+\vvertil{}{\sss{LQ}}{\sman,\tman}
                -\vvertil{}{\sss{LL}}{\sman,\tman},  
\\  
\label{form4}
\kappa_{ef}&=& 1+\vvertil{}{\sss{QQ}}{\sman,\tman}
                -\vvertil{}{\sss{LL}}{\sman,\tman}.
\end{eqnarray}
{The corresponding relations of form factors $F_{ij}$ and the $Z$ boson matrix element are:}
\begin{eqnarray}  
{\cal A}^{\sss{OLA}}_{\sss{\zb}}&=&
\ib \gspi \,e^2\,4\,\tcie\tcif\frac{\chi_{\sss{Z}}(\sman)}{\sman} 
\\\nonumber             &&
\times \Biggl\{\gadu{\mu} {\lpar 1+\gfd \rpar } \otimes
       \gadu{\mu} {\lpar 1+\gfd \rpar } \vvertil{}{\sss{LL}}{\sman,\tman}     
-4 |\qe | \stws \gadu{\mu}     
      \otimes \gadu{\mu} {\lpar 1+\gfd \rpar}\vvertil{}{\sss{QL}}{\sman,\tman} 
\\\nonumber              &&
-4 |\qf | \stws
\gadu{\mu}{\lpar 1+\gfd\rpar}\otimes\gadu{\mu}\vvertil{}{\sss{LQ}}{\sman,\tman}
+16|\qe\qf|\stwf\gadu{\mu}\otimes\gadu{\mu}
\vvertil{}{\sss{QQ}}{\sman,\tman}\Biggr\}.\qquad\quad
\label{structures}
\end{eqnarray}
So far we discussed matrix elements.
{The differential cross section for $e^+e^- \to f {\bar f}$ is:}
\bea  
\frac{d \sigma}{d \cos \vartheta} =&& \frac{\pi \alpha_{em}^2}{2s}
\Biggl\{
\left( 1+ \cos^2 \vartheta\right)\left[ K_T(\gamma) + \mathrm{Re} (\chi(s)~~ K_T(I)) + |\chi(s)|^2 ~~
K_T(Z)\right]  
\\\nonumber  
&&
+ ~ 2 \cos\vartheta \left[K_{FB}(\gamma) + \mathrm{Re} (\chi(s)~~  K_{FB}(I)) + |\chi(s)|^2 ~~
K_{FB}(Z)\right]
\Biggr\} ,
\eea
with
\bea  
\chi(s) &=& \frac{G_F}{\sqrt{2}}  ~~ \frac{M_Z^2}{8\pi\alpha} ~~  \frac{s}{s-M_Z^2+i\Gamma_ZM_Z}   .
\eea
One has to care about the choice of a constant $Z$ boson width $\gamma_Z$ or an $s$-dependent width $\Gamma_Z$
here \cite{Bardin:1988xt}.

{The effective couplings are:}
\bea  
 K_T(\gamma) =&& c_{color} ~ Q_i^2 Q_f^2 |F_{\gamma}(s)|^2
 \\\nonumber
=_{Born}&&c_{color} ~ Q_i^2 Q_f^2 ,
\\  
K_T(I) =&&  2 c_{color}  ~  |Q_iQ_f| ~   F_{\gamma}(s)^* \rho_{if}(s,t) v_i v_f 
\\\nonumber
=_{Born}&&  2 c_{color}  ~  |Q_iQ_f|v_{B,i} v_{B,f}
,
\\  
K_T(Z) =&&    c_{color}  ~ |\rho_{if}(s,t)|^2 (1+ |v_i|^2 +  |v_f|^2 + |v_{if}|^2)
\\\nonumber
=_{Born}&& c_{color}  ~ (v_{B,i}^2+a_{B,i}^2)(v_{B,f}^2+a_{B,f}^2),
\\  
K_{FB}(\gamma) =&&  0,
\\  
K_{FB}(I) =&& 2 c_{color}  ~  |Q_iQ_f| ~   F_{\gamma}(s)^* \rho_{if}(s,t)
\\\nonumber
=_{Born}&& 2 c_{color}  ~  |Q_iQ_f| ~   a_{B,i} a_{B,f}
,
\\  
K_{FB}(Z) =&&  2 c_{color}  ~|\rho_{if}(s,t)|^2  2\mathrm{Re} (v_iv_f + v_{if})
\\\nonumber
=_{Born}&&  2 c_{color}  ~ (2v_{B,i}a_{B,i})(2v_{B,f}a_{B,f}) 
.
\eea
Here, $i$ denotes the initial state and $f$ the final state. For the Drell-Yan process ${\bar q}q \to
l^+l^-$, it is $q=u,d$ and $f=l$. 
In case of polarizations, (3.32) becomes non-vanishing \cite{Bardin:1992jc}.
\\
The $c_{color}$ is the color factor, e.g. $c_{color}=3$ for initial state quarks and final state leptons.

A formula similar to (3.27) describes the special case of Bhabha scattering
\cite{Riemann:1991ga,Bardin:1999yda,Bardin:1990xe,Field:1995dk}.
The  numerical comparison  with W.~Hollik in 1990 \cite{Bardin:1990xe} seems to be the most
precise prediction for the effective Born cross-section of Bhabha scattering until today.

At the end of the subsection, we would like to emphasize that notions of form factors are not unique.
We split, for purely phenomenological reasons, the matrix element into two pieces: a photon amplitude and a
$Z$ boson  amplitude.
The calculation of the running QED coupling is technically quite different from that of the weak loop
diagrams.
So this is reasonable. 
Gauge invariance justifies it, but only if handled with care.
There are gauge dependent diagrams which mix a photon and a $Z$ boson amplitude.
So, in ZFITTER we decided to include all the corrections but the fermionic self-energy insertions, a bit
arbitrary, into the $Z$ boson amplitude.

Such a separation of photonic and weak terms is wishful also for the charged current $W$ boson mediated
amplitude.
But a gauge-invariant  separation of (virtual and real) photonic corrections from $W$ boson exchange is
impossible.
In HECTOR \cite{Bardin:1989vz,Arbuzov:1995id}, we found a way to do well-defined separations by
considering logarithmic terms and just explicitly defining some rule.
This really worked out.
Years later, when building a software for $e^+e^- \to \nu {\bar \nu} \gamma$, we could take over the weak
charged current form factor into the Monte Carlo program of S.~Jadach and Z.~Was \cite{Bardin:2001vt}.
This reaction is, for $\nu = \nu_e$, unique:
It depends both on neutral current and charged current amplitudes.\footnote{Bhabha scattering has also $s$-
and $t$-channel exchanges, but only of neutral current type.} 

Similar problems have been discussed when the ZFITTER form factors were adapted to atomic violation
measurements \cite{Bardin:2001ii}.

Note added in May 2014: 
The International Linear Collider Technical Design Report \cite{Baer:2013cma} was published June 26, 2013. The planned unprecedented 
accuracy, notably that of the Giga-$Z$ option, but also that of general weak parameter studies, goes certainly beyond the precision reached 
so far with ZFITTER and similar codes. 
A completely different application of ZFITTER is presently planned for the analysis of muon pair production data at Belle and Belle 2 
(with about  $10^9$ muon pairs) at a center-of-mass energy $\sqrt{s}=10.58$ GeV. One may study there, besides QED 
properties, the axial vector coupling of leptons \cite{Aushev:2010bq,ferber:2014-ref}. 

\subsection{Drell-Yan processes}
We went a bit into the details of a correct ansatz for the effective Born approximation in the Standard model
in $e^+e^-$-annihilation.
The situation in a Drell-Yan process is quite similar.
One may study e.g. the {  running of the weak mixing angle $\sin^2\theta_{W}^{\rm eff}(s')$} as a function
of the
scale $s'$ from a hard cross-section $\sigma_0(s')$:
\bea 
\sigma_0(s') = {\cal L}_{u}~~\sigma_0(u{\bar u}\to l^+l^-)~~ +~~ {\cal L}_{d} ~~\sigma_0(d{\bar d}\to
l^+l^-), 
\eea 
where both hard scattering cross-sections $\sigma_0(u{\bar u}\to l^+l^-)$ and $\sigma_0(d{\bar d}\to
l^+l^-)$
{depend on four complex valued, process-dependent  form factors $\rho_{ql}, \kappa_{q},
\kappa_{l},
\kappa_{ql}$} with $q=u,d$.
The $\sigma_0$ depends on $s'$, but also on the {  scattering angle $\theta$}.
Further, we have not only initial and final state photonic corrections, but also {  initial-final state
interferences}.

An elegant way to cover at least part of the complexity of all this in a modern QCD Monte Carlo program is
the following:

-- Define a photon exchange amplitude.

-- Define a Z exchange amplitude.

-- {Split the $v_{ql}$ into a $Z$-part and a photon part:}
\bea 
v_{ql} \to (v_{ql} - v_q~v_l) ~~+~~ v_q~v_l.
\eea

-- Assume a Born like structure with form factors $\rho, v_q, v_l$ and put the deviation from that structure,
which is contained in the difference $(v_{ql}-v_qv_l)$, into the photon amplitude.

In an unpublished paper of 1991 \cite{Leike:1991if}, A. Leike and T. Riemann worked out  the influence of
$Z'$-physics on the evaluation of weak form factors.
The idea of reshuffling matrix elements in form factors  was invented there and is now independently re-used
as  clever inclusion of ZFITTER's weak form factors into a Monte Carlo code, which was originally
made for 
the description of QCD corrections to Drell-Yan processes.\footnote{W.~Sakumoto, private
information and reference \cite{Aaltonen:2013wcp}.}

Evidently, once there are accurate data, one has to carefully understand how to model the correct physics
ansatz with a smaller number of parameters. 
This is under study by experimentalists presently.

\section{Real corrections, mostly due to QED \label{sec-real}}
Around 1983 we began to envisage some contribution to the description of the $Z$ boson resonance as it was
planned to be studied at LEP.
There existed several articles on electroweak radiative
corrections.
Let us mention the electroweak study by Wetzel in 1982  \cite{Wetzel:1982mh} and that by Lynn and Stuart
in 1984 \cite{Lynn:1984rk}, or the MC program MUSTRAAL by Berends, Kleiss, Jadach in 1982
\cite{Berends:1982ie}.
It was not evident to us that we might contribute some novel results, and we decided therefore to study real
photon emission first.

The Dubna group has an enormous experience in the analytical treatment of
QED corrections, first mostly applied to $t$-channel exchange processes. This was pushed by the
close contacts with Dubna experimentalists of the NA-4 collaboration at CERN.
Basics of a systematic analytical phase space handling with massive final state particles may be found in \cite{Akhundov:1976yy}.  
The subtraction method for the treatment of infra-red singularities was worked out in
1976 in a seminal paper \cite{Bardin:1976qa}.
The divergent part of the cross-section is, in simplified form, integrated over the whole phase space, and at
the same time subtracted from the exact squared matrix element.
The difference can be integrated numerically, and the isolated term is sufficiently simple for an analytical
treatment.  
In practice, this can become quite involved, see  reference \cite{Akhundov:1994my}. 

The first articles treated just photonic corrections, taking into account mass effects.
The very first one was on pure QED corrections in $e^+e^-$ annihilation, by A.~Akhundov (Baku), D.~Bardin
(Dubna), O.~Fedorenko (Petrozavodsk), T.~Riemann (Dubna):  
``Some Integrals For Exact Calculation Of QED Bremsstrahlung'', an unpublished JINR Dubna
preprint \cite{Akhundov:1984mm}, followed by \cite{Akhundov:1984mp,Fedorenko:1986hw}.
Then we extended the integration technology to experimental set-ups with $Z$ boson resonance
phenomena,
including mixing phenomena of $Z$ boson and photon.
This sounds easy, but there were several conceptual problems to be solved.
As a result, ZFITTER relies now on several versions of semi-analytical formulae with low-dimensional numerical
phase space integrations left.
At the time of LEP experiments, this was extremely useful.
For an unfolding of measured cross-sections into pseudo-observables, or for
multi-dimensional fits, the  computing time of an analysis code was absolutely decisive. 
The inclusion of certain kinematical cuts was a wish expressed by experimentalists.
Computers were not so advanced. There were no personal computers, and workstations were also not
yet on the market.
In Dubna, there were one or two terminal stations for theoreticians, and we had to queue up every day.
In Russian Winter, the terminal room (with one terminal) was a bit cold at temperatures close to zero
centigrades, because the windows did not close exactly. 
The upper left corner of the terminal screen was blind. 
Often the terminal in the theory building was blocked by Riemann, Bardin, Akhundov from 9 to 12 in the
morning.
Not everybody was amused.

In case of quark-pair production, or $Z$ or $W$ boson decays into quarks, the final state will get QCD
modifications.
The corrections are contained in so-called radiator functions.
Their implementation in ZFITTER relies on calculations by a variety of colleagues and is described in the
various ZFITTER descriptions, notably in
 references \cite{Bardin:1989di,Bardin:1992jc,Bardin:1999yda,Arbuzov:2005ma}. 
Useful representations are also e.g. \cite{Bardin:1997xq,Chetyrkin:1994js,Bardin:1999gt,Bardin:1999ak}.

The treatment of the complete set of QED corrections related to real emission of photons in ZFITTER
is quite specific. 
The higher-order corrections have been typically taken over from the literature, as it is
documented, notably in references \cite{Bardin:1999yda,Arbuzov:2005ma}.
An important example is reference \cite{Beenakker:1989km}. 
The main work had to be performed at one-loop order, plus soft photon exponentiation.
It was clear that the numerical effects will be important for the experimental analyzes.
There were several Monte-Carlo programs available, e.g.
\cite{Berends:1980yz,Berends:1982ie,Jadach:1990mz,Jadach:1991ws,Montagna:1993te,Jadach:1999vf} and the
references therein.
See also the report \cite{Altarelli:1989wu}.  
We aimed at an alternative, analytical integration of the three-dimensional photon phase space
integrals.
The necessary techniques have been developed step by step over a longer period, and originate to a large
extent from studies for deep-inelastic scattering, e.g. $lN \to lX$ \cite{Akhundov:1994my}.
In the presence of the $Z$ boson resonance in the $s$-channel, one is faced with the additional need
to perform a correct treatment of the Breit-Wigner propagator, a truly complex function.
Further, there is a mixing of photon and $Z$ boson exchange. 
This $\gamma Z$ mixing was studied e.g. in \cite{Mann:1983di,Mann:1983dv,Riemann:1988gy,Consoli:1989fg}; this
issue was
settled by a formal Dyson summation of the $\gamma, Z$ propagator matrix.
The $Z$ boson propagator with the finite width may become an issue for analytical integrations.
In squared matrix elements, we are faced with $\gamma\gamma$, $\gamma Z$ and $ZZ$ interferences.
The latter are dominating around the $Z$ boson pole, and they will contain squared $Z$ boson propagators.
To perform analytical phase space integrations with such a term inside looks difficult.
An important, simple  idea is to perform a partial fraction decomposition in order to linearize the
integrand:
\bea
\left| \frac{1}{s-M_Z^2 + i M_Z \Gamma_Z} \right|^2  
&=& \frac{-1}{2iM_Z \Gamma_Z} 
~~\left( \frac{1}{s-M_Z^2 + i M_Z \Gamma_Z}  - \frac{1}{s-M_Z^2 - i M_Z \Gamma_Z}  \right)
\\ \nonumber
&=& \frac{-1}{M_Z \Gamma_Z} \mathrm{Im}  \left( \frac{1}{s-M_Z^2 + i M_Z \Gamma_Z} \right).
\eea
At first glance this looks bizarre because the complete answer seems to carry an overall factor 
$s/(M_Z \Gamma_Z)$.
Evidently, one may use complex integration theory, so this is good.
The overall pre-factor gets divergent for  vanishing $Z$ width, but this is a technical expression of the
well-known radiative tail, so this is also good.

We tried the approach, and  calculated the complete one-loop QED corrections for the total cross-section and
the
forward-backward asymmetry around the $Z$ resonance without a cut.
The results for initial state radiation, final state radiation and the initial final state interferences were
rather compact and looked explicitly reasonably behaving.\footnote{In fact, it took us nearly half a
year of heavy fighting with SCHOONSCHIP in 1987, because we did not agree, at the $Z$ boson peak,  with
the numerics of the Monte-Carlo program MUSTRAAL
\cite{Berends:1982ie,Berends:1983mi}. The MUSTRAAL was available via CPC, and we could run it at Dubna.
The mistake was, as often, trivial, but influential. The final 5 digits agreement convinced us that our
Breit-Wigner treatment makes sense and is operational.}
The results were published as preprint in \cite{Bardin:1987hv} and refined a
bit in \cite{Bardin:1988ze}.
The paper could not be published in Nuclear Physics B because the referee found it not close enough to the
experimental set-up.
Nevertheless, it is a nice piece of work and served for many years as an important numerical etalon for
precision comparisons. 
Note added in May 2014: Such a careful comparison of the analytical predictions from \cite{Bardin:1987hv,Bardin:1988ze} with those from the 
 Monte Carlo  
generator MUSTRAAL \cite{Berends:1980yz,Berends:1982ie,Berends:1983mi} was performed in 1988 at CERN \cite{delphi:88-67-phys-30}.

As a by-product, we understood that one may calculate the photonic corrections to the initial-final state
interference of the $\gamma Z$
interference as the arithmetical means of the corrections to the $ZZ$ and $\gamma\gamma$ initial-final state
interferences:
\bea 
R_{ini-fin}(Z, Z_2) &=& \frac{1}{2}\left[R_{ini-fin}(Z,Z) +  R_{ini-fin}(Z_2, Z_2)\right],~~~Z_2 = \gamma.
\eea 
Here, $Z_2$ is a second vector boson.
For a proof see  reference \cite{Riemann:1988gy}.
This is not of utmost importance here. When we later studied QED corrections for $Z,Z'$ production with
a heavy $Z'$ boson, then we had the newly appearing initial-final state part of the $Z Z'$ interference  at
the disposal without a new calculation
\cite{Leike:1989ah,Leike:1990zp,Djouadi:1991sx,Adriani:1993ca,RIEMANNsabine:1994}.


Later we refined the techniques, and finally ZFITTER enables the calculation of 
\begin{itemize}
 \item 
exact, completely integrated one-loop photonic corrections without cuts \cite{Bardin:1988ze};
 \item 
convolution integrals for cross-sections with soft photon exponentiation \cite{Bardin:1989cw};
 \item 
the corresponding  angular distributions \cite{Bardin:1990fu};
 \item 
convolution integrals with integrated angular cuts \cite{Bardin:1990de};
 \item 
convolution integrals with integrated acceptance cuts, combined further with an acollinearity cut
\cite{Bilenky:1989zg}.
\end{itemize}
The sophisticated final state phase space treatment with cut on the acollinearity final state  fermions goes
back to G.~Passarino (1982) \cite{Passarino:1982zp} and is relatively close
to realistic 
experimental cuts for lepton final states.
The complete analytical QED corrections were worked out for this case by M. Bilenky and A. Sazonov
\cite{Bilenky:1989zg}
and became part of ZFITTER.
The truly nice paper remained unpublished, unfortunately.
Later, we recalculated these corrections for ZFITTER from the scratch (unpublished, see references 
\cite{Christova:1999gh,Christova:2000zu,Christova:1998tc,Jack:2000as}).
We performed two minor corrections and got very nice, compact formulae
for the special case of no cut for the fermion production angle \cite{Christova:1999cc}.  

Finally, all this was sufficiently close to what the experimentalists could derive  from their Monte-Carlo
simulations for a confrontation with theory.

We wrote relatively monstrous programs in Veltman's SCHOONSCHIP \cite{Strubbe:1974vj,Veltman:1991xb} to be run
at a CDC-6500 main frame at JINR Dubna.
Bardin and Fedorenko where, in parallel
with Vladimirov and Tarasov,  among the first using SCHOONSCHIP at JINR in 1976.\footnote{ 
A.~Akhundov,
D.~Bardin, L.~Bobyleva, V.~Gerdt, I.~Shidkova, W.~Lassner, V.~Rostovzev, O.~Tarasov, R.~Fedorova and 
D.~Schirkov  received in 1986 the JINR Award in Theoretical
Physics for ``Introduction, development and use of computer systems for analytical calculations at central
computers of the central computing installations of JINR''.
We are grateful to V. Gerdt for a
clarifying email exchange.}
Colleagues from Moscow came to JINR regularly in order to use the CDC-6500 main frame because comparable
computers were subject of the US embargo policy and thus not available for civil use in Soviet Union at that
time. JINR, Dubna, as an international research center,  was privileged in that respect.\footnote{We are
grateful to Andrei Kataev reminding about this fact.} 
A comprehensive review on the use of computer algebra at
JINR is \cite{Gerdt:1987}.
Later FORM \cite{Vermaseren:1991??,Vermaseren:2000nd} was invented by Jos Vermaseren and we could run it at
personal computers.
The first article typeset in latex was presumably \cite{Bardin:1990fu}, and the first article submitted to
the
hep-ph archive dates in 1994; The archive hep-ph was opened in 1992. 

At a certain moment we realized that analytical integrations are fine; but if the sensitivity to the $Z$
boson width becomes sufficiently large, then it will matter whether the width is a pure constant $\gamma_Z$ as
in a
normal Breit-Wigner function, or whether it arises from a quantum field theoretical calculation and will thus
depend on the kinematics, $\Gamma_Z(s)$.
In the latter case, it is (roughly speaking)  the imaginary part of the $Z$ boson self-energy function, which
is by itself $s$-dependent; and for initial state corrections $s'$-dependent.
The $s'$ is one of the integration variables.
We remembered that the $s$-dependence is, to a very high accuracy, just $\Gamma_Z(s) = (s/M_Z) ~
\Gamma_Z$, and
this observation enables us to change the propagators into functions with a constant width, allowing
not only a
good estimate of the different approaches, but also further-on the analytical integrations:
The differences of mass and width in the two approaches derive from the following identity
\cite{Bardin:1988xt}:\footnote{The $Z$
boson  mass shift was also discovered by a numerical study of the $Z$ boson peak in parallel to
\cite{Bardin:1988xt} in
\cite{Berends:1987bg}.}
\bea 
\frac{1}{s-M_Z^2+iM_Z\Gamma_Z(s)} &\equiv& c~~ \frac{1}{s-m_Z^2 +i m_Z \gamma_Z} ,
\eea
with
\bea
\label{eqzmt}
m_Z &=& M_Z - \frac{\Gamma_Z^2}{2M_Z} ~=~ M_Z - 34 ~{\rm MeV},
\\
\label{eqzmwt}
\gamma_Z &=& \Gamma_Z -\frac{\Gamma_Z^3}{2M_Z^2}    ~=~ \Gamma_Z - 0.934 ~{\rm MeV} \approx  \Gamma_Z -1 ~{\rm MeV} . 
\eea
Here, $M_Z = 91.1876$ GeV and $\Gamma_Z = 2.4952$ GeV  have to be chosen as the usual PDG values. 
Later we worked out an approach to a  model-independent $Z$ boson peak analysis inspired by S-matrix theory,
relying naturally on $m_Z, \gamma_Z$.
Not only for the $Z$ boson peak cross-section, but also for asymmetries.
The point here again is a proper treatment of QED corrections
\cite{Leike:1991pq,Riemann:1992gv,sriemannL3-1233,sriemannL3-1656,Adriani:1993sx,Kirsch:1994cf}.\footnote{The
corresponding software package SMATASY is supported by Martin Gr{\"u}newald.}

In fact, the idea to use  $m_Z, \gamma_Z$ instead of  $M_Z, \Gamma_Z$ was born while listening to a
talk on string theory at a conference, while reading a paper on QED
corrections with complicated phase space cuts by Passarino \cite{Passarino:1982zp}. 

The $Z$ boson parameter relations (\ref{eqzmt}) and (\ref{eqzmwt})   become essential when two-loop
electroweak corrections are determined in ZFITTER.
This is carefully described in \cite{Awramik:2006uz}, where the complete electroweak two-loop corrections to
the leptonic weak mixing angle have been calculated. See also section \ref{sec-2013}.
It is remarkable that the shift of the $Z$ boson width due to the change of scheme ($s$-dependent or
constant $Z$ boson width) amounts to 1 MeV and is larger than the corresponding shift from the genuine weak
NNLO
corrections.
Compared to the experimental error of 2.3 MeV, the shift is small.
The authors of \cite{Awramik:2006uz} did not take the correction into account because it is formally beyond
the
NNLO order and thus among the systematically neglected terms.\footnote{Ayres Freitas, private information.}
One should consider the term as an indication of the size of unknown higher-order terms.

What we describe here is about the state of real emission affairs in ZFITTER  at the end of the
nineteen-eighties.
Final state mass effect treatments were refined in \cite{Akhundov:1991qa,Arbuzov:1991pr,Jack:2000xw}.
Some additional QED corrections, due to light fermion pair emission and higher-order photonic effects, needed
for a proper treatment at LEP 2 energies were later added \cite{Arbuzov:1999uq,Arbuzov:1999cq}.
See also reference \cite{Boudjema:1996qg}. An extended discussion of higher-order QED effects in the
leading and next-to-leading logarithmic approximations can be found in 
reviews~\cite{Arbuzov:2010zz,Arbuzov:2010zza}.

Careful studies of ZFITTER physics updates originated in these years
\cite{Bardin:1998nm,Bardin:1999gt,Kobel:2000aw}.

\section{Competition and cooperation\label{sec-comp}}

\subsection{1989 - First LEP publications}
In 1989, the world changed quite a bit.
Participation at the Ringberg Workshop on LEP physics in Germany became possible \cite{Riemann:1989dj}.
The NATO supported RADCOR conference on radiative corrections and their applications to experiments in
Brighton, the first one of a series,  was open to Eastern Country physicists \cite{Bardin:1989sm,Bohm:1989ti}.
We remember the stimulating atmosphere of the 1989 LEP physics workshop at
CERN \cite{Bardin:1989sm,Bohm:1989pb}.
And LEP became operative in August 1989.
The first months were exciting.
A good knowledge of radiative corrections was needed from the very beginning, just in order to discriminate
between trivial radiative effects and New Physics.
Several unpublished ZFITTER related theory studies appeared in this period, e.g.
\cite{delphi:88-67-phys-30,Leiste:1989jh,Bardin:1989ay,Bilenky:1989ts,bardin:1988ji}. 
In \cite{bardin:1988ji}, approximate parameterizations of $O (\alpha \alpha_s)$ corrections
\cite{Djouadi:1987di} 
were derived in order to speed-up the numerics.
The Fortran routines of B. Kniehl \cite{Kniehl:1990yc} improved this later further on.
The LEP collaborations performed the first $Z$ line shape analyzes.
We were closely related to the  L3 collaboration  
\cite{Adeva:1989mn,Adeva:1989zg,Adeva:1990we,Adeva:1990yj,Adeva:1990wy,Adeva:1990vr,Adeva:1990nv,Adeva:1990ra}
and to DELPHI
\cite{Aarnio:1989tv,Aarnio:1990fe,Aarnio:1990qe,Abreu:1990qg,Abreu:1990sr}.
A review of the latter is \cite{Alekseev:2001va}.

Among the first DELPHI papers was \cite{Aarnio:1989tv}.
From the ZFITTER group, D.~Bardin and G.~Mitselmakher were DELPHI authors.
The paper quotes for the theory on the $Z$ line shape G. Burgers \cite{Burgers:1988xy} and A. Borrelli et al.
\cite{Borrelli:1989bd}.
In \cite{Abreu:1990qg}, the $Z$ line shape analysis used  the software packages ZAPPH and ZHADRO by
G. Burgers \cite{Burgers:1988xy}.
In \cite{Aarnio:1990qe}, March 1990, our papers \cite{Bardin:1989di,Bilenky:1989zg} are quoted.
And in \cite{Abreu:1990sr} the  package ZFITTER/ZBIZON with reference to the internal note DELPHI 89-71
PHYS 52 and to \cite{Bardin:1989di,Bardin:1989tq} was used.\footnote{ZBIZON is the former version of
ZFITTER. In \cite{delphi:88-67-phys-30} (7 Oct. 1988), the Fortran code had no special name yet. ZBIZON became ZFITTER, when analytical 
angular integration and common exponentiation of initial and final state radiation were introduced. In January 1991, ZFITTER version 
3.02 was released, in June 1991 v.3.05, and in September 1991 v.4.0. ZFITTER became publicly available, with user interfaces and a draft 
description, becoming later reference \cite{Bardin:1992jc}.}

A similar approach was observed in the L3 collaboration, were ZFITTER authors T. Riemann, M. Sachwitz and H.
Vogt were collaborating in 1989.
The internal note L3-001 \cite{Adeva:1989mn} quotes G. Burgers \cite{Burgers:1988xy} and
{CERN 89-08}\footnote{\url{http://cds.cern.ch/record/116932/files/CERN-89-08-V-1.pdf}}, but also our
paper \cite{Bardin:1988xt}. 
The $Z$ line shape analysis seems to be based on papers by Cahn \cite{Cahn:1986qf} and Borrelli et al.
\cite{Borrelli:1989bd}.
In \cite{Adeva:1989zg},  internal note L3-003, our package ZBIZON is quoted with reference to  L3 Internal
Note 679 as well as \cite{Bardin:1988xt} and the Zeuthen preprint  PHE 89-19 \cite{Bardin:1990fu}.
Back-up radiative corrections had been studied with ZBIZON.
 For the very $Z$ line shape fits they used
again Borrelli et al. \cite{Borrelli:1989bd}, Cahn \cite{Cahn:1986qf}, and
a paper by Jadach et al. \cite{Jadach:1991ws}, for Bhabha scattering.  
In \cite{Adeva:1989yi}, internal note L3-004,  the paper on the Z boson parameters
\cite{Bardin:1988xt} was quoted.

A bit later it became more and more common to use ZFITTER in DELPHI and L3, but also in OPAL.
While ALEPH used the package BHM/WOH by F.~Berends, M.~Martinez, W.~Hollik et al.
\cite{Hollik:1988ii,Bardin:1997xq}.
We mention these very first papers on LEP physics results because they demonstrate that there was a true
competition
of the analysis packages and our ZBIZON/ZFITTER package was accepted step by step, but not from the very
beginning.

%

\subsection{1992-2012 - LEPEWWG and global fits}

The LEP Electroweak Working Group was founded in 1993.\footnote{We  are grateful to Dorothee Schaile for
 private information.}
Soon after the first measurements at LEP the quest was expressed for combined data analyzes with a fourfold
statistics compared to a single experiment. 
Originally a group with members of the four LEP experiments, led by Jack Steinberger, investigated the
{combination of the Z line shape} \cite{Alexander:1991vi}.
In 1993 {  Dorothee  Schaile} was asked to take over the coordination of the group and she had then
already ideas on the inclusion of other electroweak observables into a combined analysis.
They called themselves the 
{LEP EWWG}\footnote{\url{http://lepewwg.web.cern.ch/LEPEWWG/}}. 
The first publicly accessible document with this name is also the
initial summary of the LEP results for the electroweak
Summer conferences in 1993, which then appeared annually 
\cite{Arnaudon:1993gv,collaborations:1993aa,Collaborations:2000aa}.
The LEP EWWG was lead by {  D. Schaile} from 1993-1996.
When she became professor in Munich, Robert Clare took over the  coordination of the LEP
EEWG.\footnote{We  are grateful to J. Mnich for a clarification.} The present chair is  Martin
Gr\"unewald.
The final paper on LEP 1 data appeared in 2005 \cite{ALEPH:2005ab}, nearly a decade after closing LEP 1 in
1996, while the analysis of LEP2 data (finalized data taking in 2000) was finished these days
\cite{Schael:2013ita}.

The ZFITTER group members, as well as the authors of other physics software packages used by the LEPEWWG are
not members of the LEPEWWG.
They are consulted in case.

\subsection{1995 -- The Electroweak Working Group Report \label{ssec-1995}}

 The work of the LEPEWWG and of the four LEP collaborations relied on ZFITTER and TOPAZ0, and also on
the BHM/WOH package, and on many other resources.
Because of this role of establishing a kind of world standard, the community felt the need of careful
numerical checks on their predictions.
One is confronted with multi-parameter problems, different calculation schemes, some freedom of
input choices, in the presence of approximations and dedicated omissions, of misunderstandings and,
sometimes, mistakes.

At a certain moment, the community has to set  benchmarks. 
The result of a year-long workshop is the  collection
"Reports of the working group on precision calculations for the Z resonance", 
edited by D.~Bardin, W.~Hollik, G.~Passarino.
It was published as a CERN Yellow Report,
 CERN 95-03 (31 March 1995), \url{http://cdsweb.cern.ch/record/280836/files/CERN-95-03.pdf}.

Part of this document is the {  "Electroweak Working Group Report"}, which was two years later submitted to
the arxive/hep-ph \cite{Bardin:1997xq}.\footnote{Now it is also available as a pdf file at CERN, in
CERN 95-03.}
{  This work is one of the basics for the successful work of the LEP Electroweak Working Group. }
It is until now one of the most important collections of Standard Model higher-order corrections
for $e^+e^-$-annihilation.

\subsection{Higher-order corrections in ZFITTER\label{ssec-higher}}

During the 1995 CERN workshop and shortly after, a lot of additional higher-order corrections were calculated
and included into ZFITTER.
We give here just a (presumably not complete) list of references and refer for any detail to the ZFITTER
descriptions:
\cite{Chetyrkin:1994js,Consoli:1989fg,Awramik:2006uz,vanderBij:1984aj,Kniehl:1988id,Kniehl:1990yc,
Barbieri:1992dq,Degrassi:1994kt,Avdeev:1994db,Chetyrkin:1995ix,Eidelman:1995ny}.
Later, further improvements were added
\cite{Schroder:2005db,Freitas:2000nv,Freitas:2000gg,Awramik:2003rn,Awramik:2004rt,Awramik:2004qv,
Awramik:2004ge,%
Awramik:2006ar,Awramik:2008gi,Baikov:2012er}.

Until now, we did not yet include into ZFITTER the existing parameterization of the rather small {\em
bosonic} two-loop weak corrections to the weak mixing angle \cite{Awramik:2006ar}.
The fermionic corrections are covered, as well as the complete weak two-loop corrections to the $W$
boson mass.
For a complete treatment of the weak two-loop corrections to the $Z$ boson width, the corrections to the form
factor  $\rho_Z$ are lacking yet.
For this reason, the quite good agreement of the higher-order {\em approximations} to $\Gamma_Z$ with the so
far known pieces of the {\em complete} two-loop result are an indication that the final answer will be close
to what we have already.

Generally speaking, we try to control about four to five digits of the predictions aiming
at such a {\em physical} theory precision.
One quote from the report \cite{Bardin:1997xq} is interesting because it sheds some light on the progress of
the so-called {\em technical} precision (precision under fixed, maybe not realistic conditions):
`` ... compare results of independent calculations. Such a comparison has been done once for {$\Delta r$,
and an agreement of up to 12 digits} (computer precision) was found [14].''
Ref. [14] was private communications of D.~Bardin, B.~Kniehl and R.~Stuart in 1992.
This has to be compared to a three digits agreement between two Bhabha cross section calculations in a
comparison, performed few years earlier in 1990 \cite{Bardin:1990xe}.
Later, in 2002, a precision of up to 12 digits was reached in practice for complete virtual one-loop
calculations, and of 5 digits with inclusion of real corrections
\cite{Fleischer:2002rn,Fleischer:2002nn,Hahn:2003ab}.

\section{ZFITTER 2013 \label{sec-2013}}
\subsection{From ZFITTER v.6.42 to ZFITTER v.6.44beta}
The most recent publicly available ZFITTER version is ZFITTER v.6.43 (17 June 2008)
\cite{Bardin:1999yda,Arbuzov:2005ma}.
It agrees with ZFITTER v.6.42 up to a correction of a non-influential typo and was
released by the ZFITTER  support group (A.~Arbuzov, M.~Awramik, M.~Czakon, A.~Freitas, M.~Gr\"unewald, K.~
M\"onig, S.~Riemann, T.~Riemann, see \url{http://zfitter.education}). The ZFITTER group was reorganized in 
February
2012 and consists now of
A.~Akhundov, A.~Arbuzov, D.~Bardin, P.~Christova, L.~Kalinovskaya, A.~Olshevksy, S.~Riemann, T.~Riemann.

\begin{figure}[ht]
\begin{center}
\includegraphics[width=10.cm]{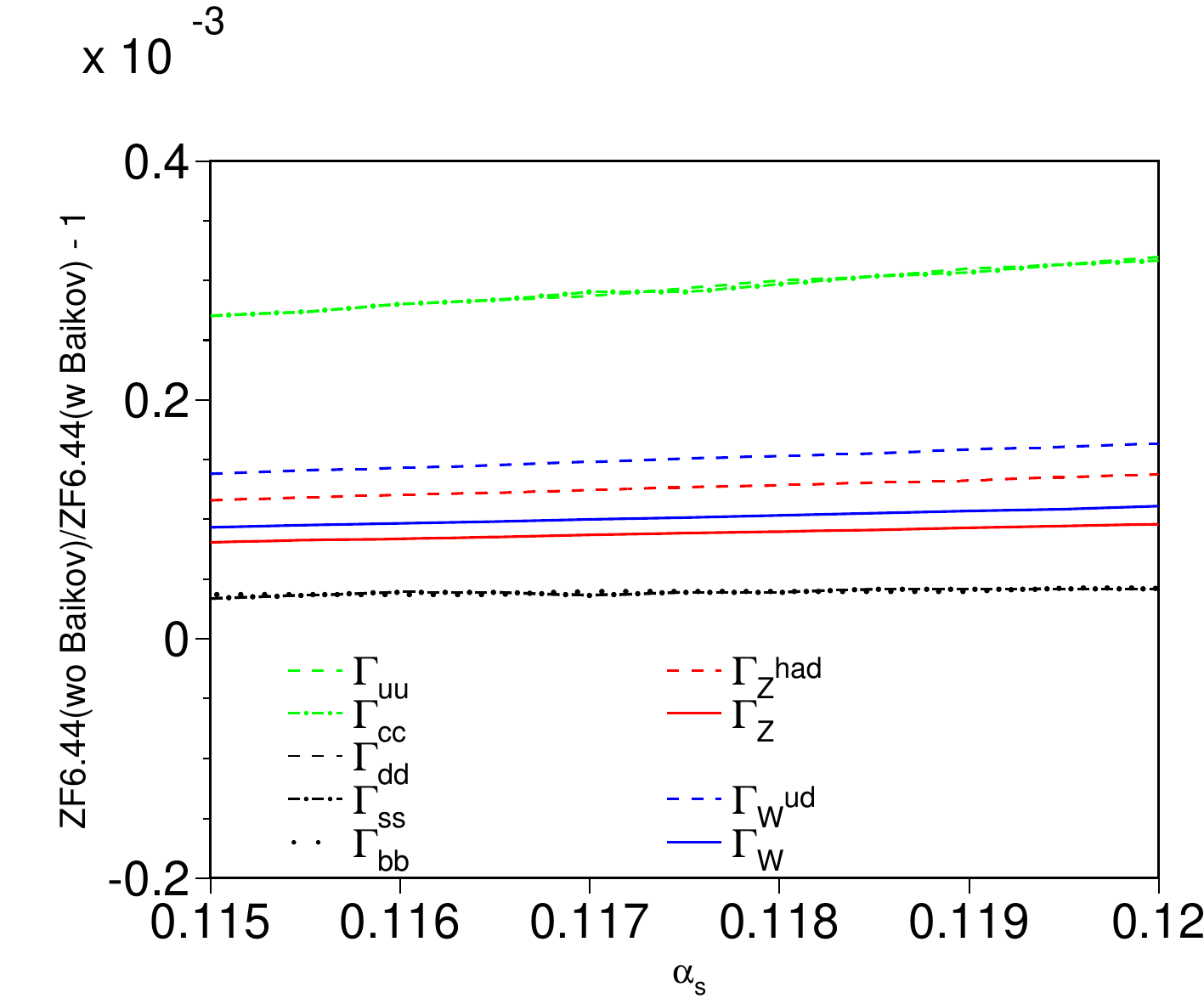}
\end{center}
 \caption[QCD NNNLO corrections on the $Z$ and $W$ widths]{\textit{
The influence of the ${\cal O}(\alpha_s^4)$ QCD corrections \cite{Baikov:2012er} on the
$W$ and $Z$ boson widths.
  \label{fbaikov2012}}}
\end{figure}

Recently, we have included into ZFITTER v.6.44beta (20 January 2013) the  final results for the ${\cal
O}(\alpha_s^4)$ QCD
corrections to the $Z$-boson and $W$-boson quarkonic partial widths and to the so-called $R$-ratio by P.
Baikov et al. \cite{Baikov:2012er}.
As may be seen from figure \ref{fbaikov2012} and from table \ref{tab-awaramik}, the numerical
shifts in the
widths amount to less than 0.3 MeV and are thus well below the experimental errors, e.g. at LEP or at an
anticipated GigaZ option of an ILC \cite{Baer:2013cma-long}.\footnote{A detailed numerical study is in
preparation.} 
A fit formula for the complete electroweak two-loop corrections to  the $W$-boson
mass \cite{Awramik:2003rn} was already included in ZFITTER v.6.42.
The final exact results for the complete electroweak two-loop corrections to  
$\sin^{2}\theta_{\rm eff}^{\rm f\bar f}$ for light fermions $f$ \cite{Awramik:2006uz} and the two-loop
electroweak fermionic corrections to $\sin^{2}\theta_{\rm eff}^{\rm b\bar b}$ \cite{Awramik:2008gi} have to be
included yet into ZFITTER. They are known to be small corrections compared to the fit formula 
\cite{Awramik:2004ge} covered in ZFITTER since v.6.42.\footnote{Note added in May 2014: A recent discussion of the status is 
\cite{freitas:LL2014}.} 
Already these corrections are small compared to the
present experimental errors for the gauge boson widths, see table \ref{tab-awaramik}. 
For the leptonic weak mixing angle, they are of the order of the experimental error: Compare the Particle
Data Group value of (\ref{sinw2-lep}) with the last row in table \ref{tab-awaramik}.
The comparison shows even a systematic deviation of the two values.
This deviation traces back to the handling of the hadronic contributions to the photonic vacuum polarization.
Changing the ZFITTER default by flag setting ALEM=2 into a variable input and setting this to  
$\Delta \alpha_\mathrm{had}^{(5)}(M_Z)=0.02750$ \cite{Jegerlehner:2011mw}, produces a shift of the
ZFITTER prediction towards the PDG
value.\footnote{Taking into account the uncertainty $\Delta \alpha_\mathrm{had}^{(5)}(M_Z)=0.02750 \pm 0.00035$ 
[\url{http://lepewwg.web.cern.ch/LEPEWWG/plots/winter2012/}], 
the corresponding predictions in table~\ref{tab-baikov2} vary:
$\Gamma_Z(\mu^+\mu^-)$ by $\pm 6.7 \times 10^{-5}$~GeV,~
$\Gamma_Z            $ by $\pm 1.2 \times 10^{-4}$~GeV,~
$\Gamma_W(l\nu)      $ by $\pm 2.2 \times 10^{-4}$~GeV,~
$\Gamma_W            $ by $\pm 2.2 \times 10^{-4}$~GeV,~
$M_W                 $ by $\pm 7.5 \times 10^{-5}$~GeV,~
$\sin^{2}\theta_{\rm eff}^{\rm lept}$ by$~\pm 5.0 \times 10^{-4} $. The latter is about the value of the
the experimental error.} 
See the changes shown in table \ref{tab-baikov2}.
Just to mention, the influence of $\Delta \alpha_\mathrm{had}^{(5)}(M_Z)$ on the Higgs mass prediction is
visualized in figure \ref{plb166fig1}, right. Here it is of minor importance, but visible.

\begin{table}[ht] 
\caption[Electroweak NNLO corrections in ZFITTER]{%
\textit{\small
ZFITTER v.6.44beta, with the input values $\alpha_s$ = 0.1184, $M_Z = 91.1876$~GeV, 
$M_H  = 125$~GeV, $m_t=173$~GeV.
The dependence on electroweak NNLO corrections is studied for IMOMS=1 (input values are 
$\alpha_{em}$, $M_Z$, $G_{\mu}$).
AMT4=4: with two-loop sub-leading corrections and re-summation recipe of [23-28] of \cite{Arbuzov:2005ma};
AMT4=5: with fermionic two-loop corrections to $M_W$ according to [29,30,32] of \cite{Arbuzov:2005ma};
AMT4=6: with complete two-loop corrections to $M_W$ [37] and fermionic two-loop corrections to 
$\sin^2\theta_W^{\rm lept,eff}$ [52] of \cite{Arbuzov:2005ma}.
IBAIKOV=0 (no $\alpha_s^4$ QCD corrections) or 
IBAIKOV=2012 \cite{Baikov:2012er}.
}
\label{tab-awaramik}
}\centering
\begin{tabular}[b]{|l|c|c|c|c||c|}
\hline
AMT4               &    4     &      5   &    6    & Diff. & Exp. Err. \\\hline\hline
\multicolumn{6}{|l|}{IBAIKOV=0}\\\hline
$\Gamma_Z(\mu^+\mu^-)$, MeV &   83.9782&   83.9748&  83.9807& 0.0059& 0.086 \\\hline
$\Gamma_Z    $, MeV         & 2494.7863& 2494.6019&2494.8688& 0.2669&   2.3 \\\hline
$\Gamma_W(l\nu)  $, MeV     &  226.3185&  226.2877& 226.2922& 0.0308&   1.9 \\\hline
$\Gamma_W    $, MeV         & 2090.3308& 2090.0465&2090.0882& 0.2843&   42  \\\hline
$M_W       $, GeV           &   80.3578&   80.3541&  80.3546& 0.0037& 0.015 \\\hline
$\sin^2\theta_\mathrm{eff}^\mathrm{lept}$
                    &  0.231722&  0.231791& 0.231670&0.000121&  0.00012    \\\hline 
\multicolumn{6}{|l|}{IBAIKOV=2012} \\\hline
$\Gamma_Z(\mu^+\mu^-), MeV$ &   83.9782&   83.9748&  83.9807& 0.0059& 0.086 \\\hline
$\Gamma_Z    $, MeV         & 2494.5591& 2494.3747&2494.6416& 0.2669& 2.3\\\hline
$\Gamma_W(l\nu)  $, MeV     &  226.3185&  226.2877& 226.2922& 0.030 & 1.9\\\hline
$\Gamma_W   $, MeV       & 2090.1117& 2089.8274&2089.8691&0.2843 &  42\\\hline
$M_W $, GeV              &   80.3578&   80.3541&  80.3546&0.0037 & 0.015\\\hline
$\sin^2\theta_\mathrm{eff}^\mathrm{lept}$    
                    &  0.231722&  0.231791& 0.231670&0.000121& 0.00012      \\\hline
\end{tabular}
\end{table}

\begin{table}[bthp]
 \caption[QCD NNNLO correcions in ZFITTER]{%
\textit{\small
IBAIKOV=0 (no $\alpha_s^4$ QCD corrections) or 
IBAIKOV=2012 \cite{Baikov:2012er}, AMT4 as described in
table \ref{tab-awaramik}. 
The difference to table \ref{tab-awaramik}:
Flag ALEM=2 is chosen with input value $\Delta
\alpha_\mathrm{had}^{(5)}(M_Z)=0.02750$. 
\label{tab-baikov2}
}}\centering
\begin{tabular}[b]{|l|c|c|c|c||c|}
\hline
AMT4               &    4     &      5   &    6    & Diff. & Exp. Err. \\\hline\hline
\multicolumn{6}{|l|}{IBAIKOV=0}\\\hline
$\Gamma_Z(\mu^+\mu^-)$, MeV  &   83.9875&   83.9839 &  83.9900 
                                                      & 0.0061 & 0.086 \\\hline
$\Gamma_Z    $, MeV      & 2495.2859& 2495.0958 &2495.3662 
                                                      & 0.2704 &   2.3 \\\hline
$\Gamma_W(l\nu)  $, MeV  &  226.4020&  226.3703 & 226.3745 
                                                      & 0.0317 &   1.9 \\\hline
$\Gamma_W    $, MeV      & 2091.1020& 2090.8092 &2090.8474 
                                                      & 0.2928 &   42  \\\hline
$M_W       $, GeV        &   80.3677&   80.3639 &  80.3644 
                                                      & 0.0038 & 0.015 \\\hline
$\sin^2\theta_{\mathrm{eff}}^{\mathrm{lept}}$
                    & 0.231532 & 0.231603  & 0.231481 
                                                      & 0.000122 &0.00012\\\hline 
\multicolumn{6}{|l|}{IBAIKOV=2012}\\\hline
$\Gamma_Z(\mu^+\mu^-)$, MeV  &   83.9875&   83.9839 &   83.9900 & 0.0061 & 0.086 \\\hline
$\Gamma_Z    $, MeV      & 2495.0586& 2494.8685 & 2495.1389 & 0.2704 & 2.3\\\hline
$\Gamma_W(l\nu)  $, MeV  &  226.4020&  226.3703 &  226.3745 & 0.0317 & 1.9\\\hline
$\Gamma_W   $, MeV       & 2090.8828& 2090.5901 & 2090.6283 & 0.2927 &  42\\\hline
$M_W $, GeV              &   80.3677&   80.3639 &   80.3644 & 0.0038 & 0.015\\\hline
$\sin^2\theta_{\mathrm{eff}}^{\mathrm{lept}}$    
                    & 0.231532 & 0.231603  & 0.231481  & 0.000122 &0.00012\\\hline
\end{tabular}

\end{table}

Presently, there are controversial positions concerning ZFITTER's `conditions of use' and the ZFITTER
software
license \url{http://cpc.cs.qub.ac.uk/licence/licence.html} granted to the authors by Elsevier's Computer
Physics Communications Program Library - Programs in Physics \& Physical Chemistry. 
For some details see \url{http://zfitter.education}.
Until the issue is settled, actualized versions of ZFITTER will stay at the beta level and cannot be
released. 

\bigskip

Sooner or later, the LHC is becoming a precision tool and the community feels some steady need of
high-precision Standard Model predictions.
Both for use in global fits and for specific cross-section predictions, notably of Drell-Yan processes via the
$Z$ resonance.
This need would become even more pronounced if the ILC project would be substantialized
\cite{Baer:2013cma-long}.

\medskip

Regrettably, we see today no alternative project to ZFITTER in the field of precision Standard Model
predictions.
In the mid-nineteen nineties there were three competing (and cooperating) projects at the disposal
\cite{Bardin:1997xq}: BHM/WOH by W.~Hollik et al., TOPAZ0 by G. Passarino et al., and ZFITTER by D.~Bardin et
al.
BHM/WOH was available on request, and the latter two are publicly available. 
To our knowledge, updating and user support have been minimized for 
{TOPAZ0}\footnote{\url{http://personalpages.to.infn.it/~giampier/topaz0.html}}
 and BHM/WOH \cite{Hollik:1988ii}.

\subsection{A comment on the Gfitter project\label{sec-gfitter}}

Sometimes the Gfitter project, introduced at the webpage  \url{http://gfitter.desy.de}, is considered as an independent implementation of 
Standard Model predictions for some pseudo-observables, and as a true scientific alternative to ZFITTER (for these
pseudo-observables).
We do not share this opinion  
and would like to give a short, clarifying comment on the situation.

\medskip

The Gfitter project was started in Summer 2006 and presented to the public in December 2007, at the
kick-off meeting of the German ``Helmholtz Alliance for Physics at the Terascale'', see the slides at
\url{http://indico.desy.de/materialDisplay.py?contribId=36&sessionId=15&materialId=1&confId=477}.
Until August 2012, the Gfitter software was proprietary, but by private
information\footnote{Private information from and documentation by A. Akhundov, S. Riemann, T. Riemann, March
to May 2011, \url{http://zfitter.com}.
Further, a German ombuds person's report announces in July 2012 : ``A diploma
 thesis derives from ZFITTER in the sense that 8200 lines have been taken over by copying from 
ZFITTER.''
In the thesis work the kernel of the Gfitter/GSM software was written (in collaboration with others), and
its text delivered basic building blocks for the so-called main article on Gfitter \cite{Flacher:2008zq0}.
A third evidence for the confidential take-overs may be found in the unpublished version of
Gfitter of
July 2011, where about 100 to 200 identities are denoted, by the Gfitter/GSM authors, to originate from
ZFITTER v.6.42. On occasion of the Erratum \cite{Flacher:2008zqErr} to \cite{Flacher:2008zq0}, ZFITTER authors
wrote a letter to the Editorial Board of "European Physical Journal C" 
(14 September 2012), \url{http://zfitter.com/letter-to-the-epjc-editors.pdf}.
Note added in May 2014: For further evidence, see 
also \url{http://zfitter-gfitter.desy.de}
and \url{http://zfitter.education/2014-05-05-zfitter-zu-desywebgfitter20140404-de-long.pdf}.}
it
became known that the Standard Model library of Gfitter, Gfitter/GSM, was relying on the FORTRAN
package
ZFITTER v.6.42 and was created to a large extent by copy-paste-adapt. 
Without any  proper citation in the academic meaning of the word. 


There are several versions of the program Gfitter.

\begin{itemize}
 \item 
Gfitter/GSM (Summer 2006 - July 2011) is unpublished. It relies essentially and directly on the Standard
Model
implementation of the ZFITTER software.
On top of that, Gfitter/GSM contains few add-ons.
The \emph{electroweak add-on} of Gfitter/GSM, compared to ZFITTER v.6.42, are the bosonic
two-loop corrections to the weak mixing angle in Awramik~et~al.~\cite{Awramik:2006uz}. They are small;
see
the discussion above.
The complete two-loop parameterizations in \cite{Awramik:2006uz}, in turn, have been made with use of
ZFITTER v.6.42.
As a consequence, it is formally correct to quote for the parameterization only \cite{Awramik:2006uz},
but one should have in mind that there is inside also ZFITTER numerics.
There is also a \emph{QCD add-on} of Gfitter/GSM (2011), compared to ZFITTER v.6.42 (2006), based on
\cite{Baikov:2008jh}.
It is also numerically small (see the discussion above) and is implemented in ZFITTER v.6.44beta.

Use of this Gfitter version deserves a citation not only of \cite{Flacher:2008zq0}, but also of
\cite{Bardin:1999yda,Arbuzov:2005ma}, for using ZFITTER v.6.42, according to ZFITTER's CPC license. 

\item 

Gfitter/GSM (August 2011 till August 2012) is unpublished. According to the authors, the program relies
on a proprietary
implementation of Standard
Model corrections which are
based on a
parameterization tracing back to Cho et al. (1999) \cite{Cho:1999km}, which in turn is based on an electroweak
one-loop calculation published in 1994 \cite{Hagiwara:1994pw}. 
There have been made improvements later, 
and in a recent article by Cho et al. (2011) \cite{Cho:2011rk} the authors confirm the 
reliability of their parametrization by comparing them with ZFITTER  v.6.42 predictions.
These parameterizations are  used in Gfitter further on,
and overlaid with the most recent higher-order corrections mentioned.

\item 

Gfitter\_1.0 has been released publicly in September 2012. The Standard Model
library Gfitter\_1.0/gew relies presumably on the same parameterizations as Gfitter/GSM (2011).
\end{itemize}

The different versions of Gfitter rely in one way or the other on ZFITTER v.6.42.
We further remark that without studying the numerical reliability of Gfitter,  to four or five significant
digits, 
the scientific value of the inclusion of NNLO weak and  $\alpha_s^4$ QCD corrections 
in Gfitter remains questionable.
According to our standards, Gfitter simulates Standard Model predictions with unknown precision.
It is a nice tool for the production of figures for the
illustration of Standard Model physics.
Possibly it is useful for studies beyond the Standard Model.


\section{Conclusions}

A talk on history and features of the ZFITTER project was presented at LL2012, the eleventh
``Loops and Legs'' meeting.
Its title was
``ZFITTER - 20 years after''.
\footnote{This text is an extended version of the talk. For the slides see
\url{https:%
//indico.desy.de/getFile.py/access?contribId=29&sessionId=10&resId=0&materialId=slides&confId=4362}.
The contribution to the proceedings of LL2012 in ``Proceedings of Science'' (PoS),
by A. Akhundov et al., did not appear. See for conference 
\url{http://pos.sissa.it/cgi-bin/reader/conf.cgi?confid=151} and for contribution
  \url{http://pos.sissa.it/archive/conferences/151/036/LL2012_036.pdf}.}
The ``Loops and Legs'' conference was founded by the
Zeuthen  Theory Group in 1992 when the Zeuthen Institute for High
Energy Physics of the (then already former) East German Academy of Sciences became part of DESY.
We are glad that this conference attracts since then regularly colleagues who contribute to the progress in
the field. 
A field, comprising both the branch of applied calculations and that of  development of new
theoretical methods. 

ZFITTER is certainly one of the oldest source-open software projects in elementary particle physics with a
permanent support.
It comprises practically all the theoretical knowledge of  relevance for a precise description of the $Z$
boson resonance in $e^+e^-$ annihilation and for $Z$ boson's part in global fits in the Standard Model
\cite{Riemann:2010zz}. 
Obviously, today one would create such a project quite differently. 
We can only encourage our colleagues to try. 
Complex projects need (independent) duplication.
As concerning the ZFITTER code, it is certainly of interest as a benchmark for SM calculations
in the LEP energy range. In particular it is used for cross checks in development of new codes,
see {e.g.}~\cite{Ciuchini:2013pca}.

Higher-order quantum field theoretical predictions face another problem:
The solutions become so lengthy and complex that the idea of source-open software is, in practice, 
no longer a realistic option.
This happens already with the ${\cal O}(\alpha_s^4)$ QCD corrections and the complete NNLO 
weak corrections in ZFITTER. 
They are mere parameterizations of huge, partly unpublished  expressions.

The LEP/SLC era  gave the scientific community unprecedented precision in several fundamental 
quantities like $M_Z$, $\Gamma_Z$, the effective weak mixing angle $\sin^2\theta_{W}^{\rm eff}$, 
the number of light neutrino flavors $N_{\nu}$.
Of comparable importance is the experimental confirmation of the Standard Model, a gauge theory with
spontaneous symmetry breaking, as a consistent quantum field theory, with inclusion of higher orders 
of perturbation theory.

At the present moment, the Standard Model remains being the most successfully theory in description
fundamental interactions. In fact, it possesses a huge predictive power and provides very
accurate predictions for many observables which appear to be in agreement with experimental data.
We see that also in post-LEP experiments at high-energy colliders like Tevatron and LHC
as well as in high-precision low-energy experiments like searches for rare decays.
Even so that we hardly believe that the Standard Model is the true theory of everything, 
it will certainly remain to be our working high-energy physics tool in the most relevant 
energy domain. 

\medskip

We are proud that we are being contributing to the establishment of the Standard Model.

\section*{Acknowledgments}
\addcontentsline{toc}{section}{Acknowledgments}

We would like to thank 
S.~Alekhin,
J.~Bl{\"u}mlein,
K.~Chetyrkin,
A.~Freitas,
S.~Moch
for helpful discussions, and 
J.~Mnich for a careful reading of the article.
D.~Bardin and L.~Kalinovskaya interacted with us when figures and tables were produced.
The authors list of this resume of ZFITTER might be much longer, as may be seen from the  list
of authors of ZFITTER and from the
list of references of the present text. 
We are truly thankful to our co-authors, users, competitors for many years of common scientific
work. 
Our friendship is alive, while times they were are changing.

This work is supported by the European Initial Training Network LHCPHENOnet
PITN-GA-2010-264564. 
A.B.A. is grateful for support to the Dynasty foundation.

Note added in May 2014: The ZFITTER project comprises about 30 person years of theoretical research. In prices of senior staff researchers 
in 2014 in Europe, this investment corresponds to about 2.2 millions of Euros. We would like to thank the many organizations who made the 
project possible, notably the Joint Institute for Nuclear Research (JINR), Dubna, Russia. 

\medskip

\clearpage

 \addcontentsline{toc}{section}{References}
\small 

\providecommand{\href}[2]{#2}\begingroup\raggedright\endgroup

\end{document}